\documentclass[amsmath,amssymb,twocolumn,pre,floatfix]{revtex4}

\usepackage{siunitx}

\usepackage{color}
\usepackage{graphicx}
\usepackage{dcolumn}
\usepackage{times}
\usepackage{soul}
\usepackage{hyperref}
\usepackage{tabularx}
\newcolumntype{b}{>{\hsize=.12\textwidth}X}
\newcolumntype{s}{>{\hsize=.15\textwidth}X}

\usepackage{afterpage}

\begin{document}

\title{Shape multistability in flexible tubular crystals through interactions of mobile dislocations} 

\author{Andrei Zakharov and Daniel A.~Beller}
    \email[Correspondence email address: ]{azakharov@ucmerced.edu, dbeller@ucmerced.edu}
    \affiliation{Department of Physics, University of California, Merced, Merced, CA 95343, USA}


\begin{abstract}
We study avenues to shape multistability and shape-morphing in flexible crystalline membranes of cylindrical topology, enabled by glide mobility of dislocations. Using computational modeling, we obtain states of mechanical equilibrium presenting a wide variety of tubular crystal deformation geometries, due to an interplay of effective defect interactions with out-of-tangent-plane deformations that reorient the tube axis. Importantly, this interplay often stabilizes defect configurations quite distinct from those predicted for a two-dimensional crystal confined to the surface of a rigid cylinder. We find that relative and absolute stability of competing states depend strongly on control parameters such as bending rigidity, applied stress, and spontaneous curvature. 
Using stable dislocation pair arrangements as building blocks, we demonstrate that targeted macroscopic three-dimensional conformations of thin crystalline tubes can be programmed by imposing certain sparse patterns of defects. Our findings reveal a broad design space for controllable and reconfigurable colloidal tube geometries, with potential relevance also to carbon nanotubes and microtubules. 
\end{abstract}
 
 \maketitle

\section{Introduction} 

Diverse biological and synthetic systems at a range of scales are self-organized in  ordered two-dimensional assemblies of cylindrical topology, including single-walled carbon nanotubes (SWCNTs) \cite{thess1996crystalline}, filamentous viral capsids \cite{klug1999tobacco}, microtubules (MTs) \cite{nogales2001structural} and colloidal systems \cite{tymczenko2008colloidal}. Such \textit{tubular crystals} frequently have circumferences of order only ten times the interparticle spacing. This has the important consequence of restricting the orientations of the crystal axes, which trace out helical paths called \textit{parastichies}, to a discrete set of possible angles with the tube axis. The number of distinct parastichies defines a pair of integer \textit{parastichy numbers}, which index the possible crystalline tessellations of the cylinder, and which for SWCNTs determine  the nanotube's electrical conductivity  \cite{yakobson1998mechanical}. The ``parastichy'' terminology arises from an intriguing connection with the botanical study of phyllotaxis, which examines plant structures with repeating patterns that follow parallel helices (or spirals); examples include the arrangements of seeds on a pine cone, scales on a pineapple, or leaves on a stem  \cite{adler1997history, pennybacker2015phyllotaxis}. 
    
Along with the importance of tubular crystals to molecular biology and the study of 2D solids, there is a growing interest in tubular crystals among soft matter physicists, due to the potential for exploiting phyllotaxis as a self-organization principle for colloidal particles or nanoparticles, and thus for creating assemblies of controllable helical pitch and chirality \cite{li2005fabrication,tymczenko2008colloidal,sadoc2012phyllotaxis,avan2020self,li2019self}. Higher-scale organization can occur through the coexistence of distinct phyllotactic tessellations on the same tube \cite{lohr2010helical}. Topological defects are central to this higher-scale organization: a change in parastichy numbers requires one or more dislocation defects at the boundary between domains \cite{harris1980tubular}. In SWCNTs, the analogous Stone-Wales defects in the honeycomb lattice of graphene are of great interest for their influence on plastic deformations and electrical conductivity \cite{yakobson1998mechanical}. Similarly, the observation of microtubules with varying protofilament number (i.e.\ circumference) along their length implies the presence of dislocations in the rhombic packing of tubulin proteins \cite{chretien1992lattice}. 
    
Much of the recent work on frustrated phyllotactic self-organization has focused on particles constrained to lie in  a fixed cylindrical surface \cite{mughal2011phyllotactic,wood2013self,mughal2014theory,fu2016hard}. That version of the tubular crystal is realized in recent experiments on colloids in capillary confinement \cite{moon2004fabrication, li2005fabrication,tymczenko2008colloidal} as well as a macroscopic ``magnetic cactus'' model \cite{nisoli2009static,nisoli2010annealing}, and it gives rise to a rich variety of phenomena such as oblique (rhombic) lattices, helical faults known as line slips, and individual dislocations behaving analogously to infinite grain boundaries \cite{Amir13}. 

However, in order to  design colloidal analogues of MTs and SWCNTs, we must examine a different version of the problem, namely the \textit{freestanding} tubular crystal. Here, the particles are not constrained to any fixed surface; instead, the tubular surface emerges as the set of mechanical equilibrium positions of particles whose bond network has the topology of a tube \cite{BellerPRE16}. The ability of the tube shape to adapt locally and dynamically removes the source of frustration, namely the fixed circumference, that underlies the rich defect phenomenology of the fixed-cylinder crystals. On the other hand, this geometrical adaptability offers potential routes to stabilizing nontrivial tube shapes in the presence of defects, a possibility that has remained mostly unexplored. 
    
In this work we demonstrate numerically that freestanding tubular crystals  possess controllable and composable mechanically multistable geometries, enabled by rearrangements of dislocations through glide mobility. This effect, which we refer to as ``dislocation-mediated shape multistability,'' operates through a newly identified co-stabilization of kinks in the tube axis with defect configurations that would be unstable on the fixed cylinder or the plane. While such kinks can also be formed with isolated disclinations \cite{dunlap1994constraints,bowick2009two}, our focus on dislocation pairs emphasizes mechanical shape-reconfigurability through dislocation glide moves, which are purely local disruptions in the lattice.  
    
Using a minimal model of a tubular crystalline membrane, we show not only  that a tubular crystal may possess distinct, mechanically stable geometries, but also that it is possible to controllably switch between these competing states. Such switching can be achieved through quantitative changes in material properties, which in principle might be accomplished \textit{in situ} by varying the temperature, or through application of external bending forces in a reversible shape-memory effect. 
{Furthermore,} by repeating certain stable dislocation motifs along the length of a tubular crystal, we  show how multiple-kink structures approximating arbitrary curves can be targeted as equilibrium geometries, {offering proofs of principle for large-scale shape manipulations of tubes into bent and helical structures}. 
    
The important distinction between freestanding and fixed-surface crystals is well-known for the spherical topology,  providing  the difference between the scars of ``spherical crystallography'' \cite{bausch2003grain} on rigid spheres and buckled crystalline shells resembling viral capsids \cite{lidmar2003virus} in flexible membranes. More generally, ordered soft matter confined to rigid, curved surfaces tends to relax the stresses imposed by Gaussian curvature through pair-nucleation of topological defects \cite{bowick2009two, vitelli2006crystallography,ellis2018curvature}; conversely, freestanding crystalline membranes can spontaneously adopt buckled geometries in the presence of  defects such as dislocations and disclinations  \cite{seung1988defects,zhang2014defects}. An analogous effect in nematic elastomer sheets allows for targeted shape transitions  in the vicinity of a topological defect \cite{modes2011gaussian,zakharov2015reshaping}, where non-zero Gaussian curvature arises to relax the elastic stresses.
    
At the same time, it is worth emphasizing key differences between tubular and spherical crystals.  For spheres, topology demands a net excess disclination charge of +2, whereas tubes (even if closed as a torus) have  no topologically required defects. In addition, the Gaussian curvature of surfaces such as spheres and tori can promote lines of dislocations known as scars \cite{bausch2003grain} and stabilization of excess unbound disclinations \cite{giomi2008elastic}. In contrast, the Gaussian curvature in a perfect cylinder is  everywhere zero. Rather, defects in a tubular crystal are intimately related with the discretization of crystal axes orientations: a change of {parastichy numbers,} whether occurring spontaneously or to {mediate} plastic deformation, necessarily requires an intervening dislocation.
    
We study the emergent interplay of dislocation interactions and surface deformation geometries by modeling the  crystalline membrane as a triangular-lattice network of harmonic spring bonds with a bending rigidity, and with the overall topology of a tube. We assume that dislocations glide freely into energy-minimizing configurations, with a rapid relaxation of the crystal's elastic energy between glide steps, and with a small, finite temperature able to overcome the Peierls barrier, which we ignore \cite{hirth1983theory,BellerPRE16}. We prohibit climb motion for simplicity, with the justification that climb requires an exchange of mass with the surrounding medium along with breaking and forming of multiple bonds, a process typically much slower than glide relaxation \cite{hirth1983theory}. By this means we calculate effective energy landscapes for dislocations interacting on a tubular crystal whose surface deformations respond strongly to changes in defect position, creating  multistable energy landscapes in both defect configuration and tube shape. 
	
Our approach extends the methodology  widely used to create soft elastic actuators by encoding locally preferred membrane geometries {in the in-surface order} to develop target three-dimensional morphologies \cite{klein2007shaping,gladman2016biomimetic,aharoni2018universal}.     {Our} findings  extend our understanding of the mechanism behind the large deformations in tubular crystals occurring in nature \cite{brangwynne2007force}. Defects in the tubulin lattice can significantly alter the mesoscale shape of the MT, and can provide a means of plastic deformation to accommodate external stress without breaking \cite{janke2017causes} or folding \cite{hunyadi2007microtubule} or even reinforce the structure \cite{witten2007stress}.

\section{Minimal model of a freestanding tubular crystal}
	
	We model the tubular crystal as a network of harmonic spring bonds $l_{ij}$, connecting massless nodes $i,j$, with a bending rigidity $\kappa$, that penalizes deviations of the mesh from the preferred curvature.  The network is six-coordinated, making a triangular lattice, everywhere except at elementary dislocations, where a five-coordinated positive disclination is adjacent to a seven-coordinated negative disclination. The discrete elastic energy associated with the stretching and bending of the mesh is defined as
	\begin{align}
		\mathcal{F}^{\textrm{d}} &= \frac{\epsilon}{2}\sum_\mathrm{bonds} \left(l_{ij} - a\right)^2 + \frac{\kappa}{2} \sum_\mathrm{nodes} (4H_i^2-2 K_i).
		\label{Fe}
	\end{align}

    The first term assigns a spring constant $\epsilon$ to deviations of bond length $l_{ij}$ from rest length $a$. The second term assigns the bending rigidity $\kappa$ to curvature distortions calculated from the discrete mean curvature $H_i$ and discrete Gaussian curvature $K_i$ at each node $i$ (see Appendix A for details).
    	
    This free energy could represent a discretization of a continuum, isotropic elastic free energy (Eq.~B1 in Appendix B) for a membrane with Young's modulus $Y=2\epsilon/\sqrt{3}$, Poisson ratio $\nu=1/3$, and Lam\'e coefficients $\lambda=\mu=\sqrt{3}\epsilon/4$. However, throughout this work we treat the nodes of the mesh as the inherently discrete particles that make up the phyllotactic tessellation. From Eq.~\ref{Fe}, along with the proportionality of $\epsilon$ to $Y$ and the tube radius $R$, we obtain two important dimensionless parameters: the F{\"o}ppl-von~K{\'a}rm{\'a}n number $\gamma \equiv YR^2/\kappa$, and the reduced bending rigidity $\tilde \kappa \equiv \kappa / (Y a^2)$.
    	\begin{figure}[t]
    		\centering
    		\includegraphics[width=0.5\textwidth]{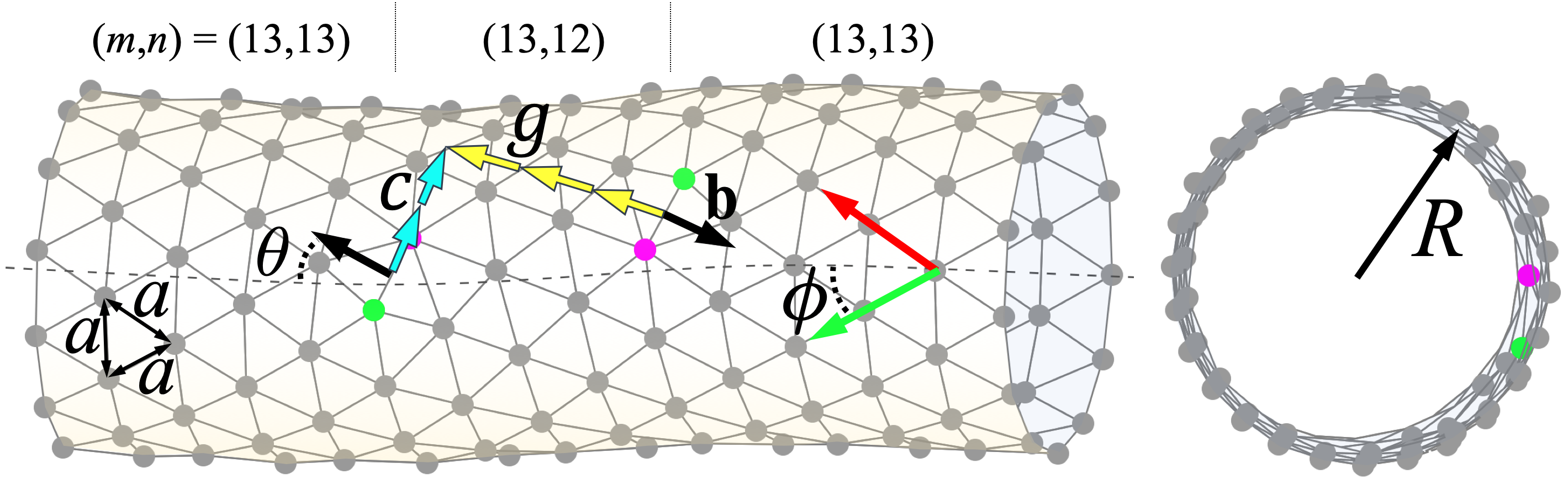}
    		\caption{ A crystal with tubular geometry of radius $R$, preferred lattice spacing $a$, and helical angle $\phi$ that defines deviation of the steepest left-handed helix (along the green arrow) from the tube axis (the dashed line). Light gray spheres each share a bond with six neighbors, whereas green and magenta spheres mark disclination sites with five or seven neighbors, respectively. Each 5-7 disclination pair forms a dislocation,  characterized by the Burgers vector $\mathbf{b}$ (black arrows) at angle $\theta$ to the tube axis. Each  dislocation changes the parastichy numbers $(m,n)$ by  $\Delta m = \pm 1$ and/or $\Delta n = \pm 1$. The tube is flexible, and thus a defect creates deformations associated with change in the tube radius and a small reorientation in the tube axis. The two dislocations are separated by $g$ glide steps (yellow arrows) and $c$ climb steps (cyan arrows), respectively along and perpendicular to $\mathbf{b}$. The two panels show the same crystal from side and top views. }
    		\label{Fig:Geometry}
    	\end{figure}
    The possible triangular lattice tessellations of a (quasi)-cylindrical surface  can be conveniently {indexed} using the botanically-inspired parastichy numbers, a pair of {integers} 
    $(m,n)$ defining the number of distinct helices of particles in the steepest right-handed and steepest left-handed families, respectively, with the former family making an 
    angle $\phi$ with the tube axis (Fig.~\ref{Fig:Geometry}). Conventionally a third parastichy number $|m-n|$, determined by the other two, is often included \cite{erickson1973tubular}. For a pristine (defect-free) tubular crystal, the radius in the limit of small $\kappa$ is uniform throughout the tube and is given by 
         \begin{align}
         	R \approx R_0(m,n) = (a/2\pi) \sqrt{m^2+n^2-mn}.
         	\label{Rmn} 
         \end{align} 

    However, for tubular crystals with  defects, the tube radius is non-uniform, and we calculate a local $R_i$ at each node $i$ as the distance from a computed centerline (see Appendix E for details).

     We consider elementary dislocations, each characterized by a Burgers vector $\mathbf{b}$, which has  length $a$ and is oriented orthogonally to the bond connecting the 5-7 disclination pair \cite{landau_elasticity} and parallel to one of the three {parastichies passing through the dislocation site} 
    The parastichy numbers $(m,n)$ thus define six possible orientations for $\mathbf{b}$, each making  a fixed angle $\theta= \phi + j\cdot 2\pi /6$, $j\in \mathbb{Z}$ with the tube axis. 
 
    We examine stability of dislocation positions with respect to glide motion, which carries a dislocation {parallel to}  $\pm \mathbf{b}$ along {one of the helical parastichies} in discrete steps of size $a$. Dislocations with collinear Burgers vectors may sit on different parastichies,   separated by climb motion normal to $\mathbf{b}$. Thus, in general, we can use the number of glide steps $g$ and climb steps $c$ separating the two parallel dislocations to parametrize axial separation $x= c a(\sqrt{3}/2) \sin \theta+ g a \cos \theta$ and azimuthal separation $y= -c a (\sqrt{3}/2) \cos \theta+ g a \sin \theta$  for a $(\mathbf{b}$, $-\mathbf{b})$ defect pair with orientations $(\theta$, $\theta+\pi)$ (Fig.~\ref{Fig:Geometry}).
     
    In this work we choose a few representative $(m,n)$ tessellations to explore in detail. Motivated by studies of microtubules, which are tubular rhombic crystals of tubulin proteins most often comprising 13 protofilaments \cite{schaedel2019lattice}, we consider defects in tubular crystals with similar ratios of circumference to bond length: the zigzag achiral configuration $(m,n)=(13,13)$, and the  armchair achiral configuration $(m,n)=(14,7)$. Our methods and qualitative findings can be extended to tubular crystals with other chirality, radius, or even lattice symmetry \cite{plummer2020buckling}, presenting a broad design space for future investigations. In this work, we use our representative tubular crystals as test beds for a systematic construction of stable tube deformation geometries built up from dislocation pair interactions.

\section{Kinked tubes and multistability in defect pair-interactions}

    We begin by examining stable configurations containing two interacting dislocations; these configurations will then be used as basic building blocks for more complex geometries. For simplicity we choose the two Burgers vectors to be antiparallel to each other, $\mathbf{b}'=-\mathbf{b}$, so that a Burgers circuit enclosing both defects measures a vanishing total Burgers vector. The effective interaction energy of the dislocation pair depends on the separation $x,y$ along the cylinder axis and the azimuthal direction, respectively, and on the Burgers vector inclination angles $\theta$, $\theta'=\theta+\pi$  relative  the cylinder axis. Dislocations initialized on the same glide parastichy may pair-annihilate on reaching zero separation, $(x,y)=(0,0)$, leaving a defect-free lattice at the energy minimum. This cannot occur for dislocations on different glide parastichies because climb would be needed to achieve zero separation. 
    
    While an analytical treatment of defects in freestanding tubular crystals is beyond the scope of this work, we turn to analytical predictions for the fixed-cylinder case as a point of comparison \cite{Amir13}. For general separations of a $\mathbf{b},-\mathbf{b}$ dislocation pair on a fixed-cylinder tubular crystal, integration of the defects' stresses gives an  effective interaction energy \cite{Amir13,BellerPRE16}
	\begin{align}
		\mathcal{F}_{\mathrm{int}}(x,y,\theta) =& \frac{A a^2}{2} \biggl\{ \ln[\cosh \tilde x - \cos \tilde y]  \nonumber \\
		& \qquad \quad   + \tilde x\left[\frac{\sinh \tilde x \cos(2\theta)+\sin \tilde y \sin (2\theta)}{\cos \tilde y - \cosh \tilde x}\right]  \biggr\} ,
		\label{Fan}
	\end{align}
	where $\tilde x \equiv x/R, \tilde y \equiv y/R$ are dimensionless quantities. Although this expression comes from a continuum elasticity calculation, it was found to agree well with simulations of fairly small cylindrical crystals \cite{Amir13}. However, the fixed-cylinder case described by Eq.~\ref{Fan} has strictly zero Gaussian curvature and prohibits  changes in tube radius that are energetically preferred  when the phyllotactic indices change on either side of a dislocation, in accordance with Eq.~\ref{Rmn}. In contrast, as we show below, a freestanding tubular crystal partially screens this interaction energy by kinking at the dislocation sites, resulting in defect pair-interaction landscapes that can differ qualitatively from the fixed-cylinder prediction of Eq.~\ref{Fan}. It is also noteworthy that Eq.~\ref{Fan} is even in $x$, a symmetry that we will see is broken by the freestanding tube. We can expect to find significant deviations from Eq.~\ref{Fan} at defect separations on the order of $\sim R/\gamma^{1/4}$, which is the length scale over which surface deformations from one dislocation can influence the other dislocation \cite{BellerPRE16}. 

	\begin{figure}[t]
		\centering
		\setlength{\tabcolsep}{-0.1em}
		\begin{tabular}{cc}
			(a)&(b)\\
			 \includegraphics[width=0.25\textwidth]{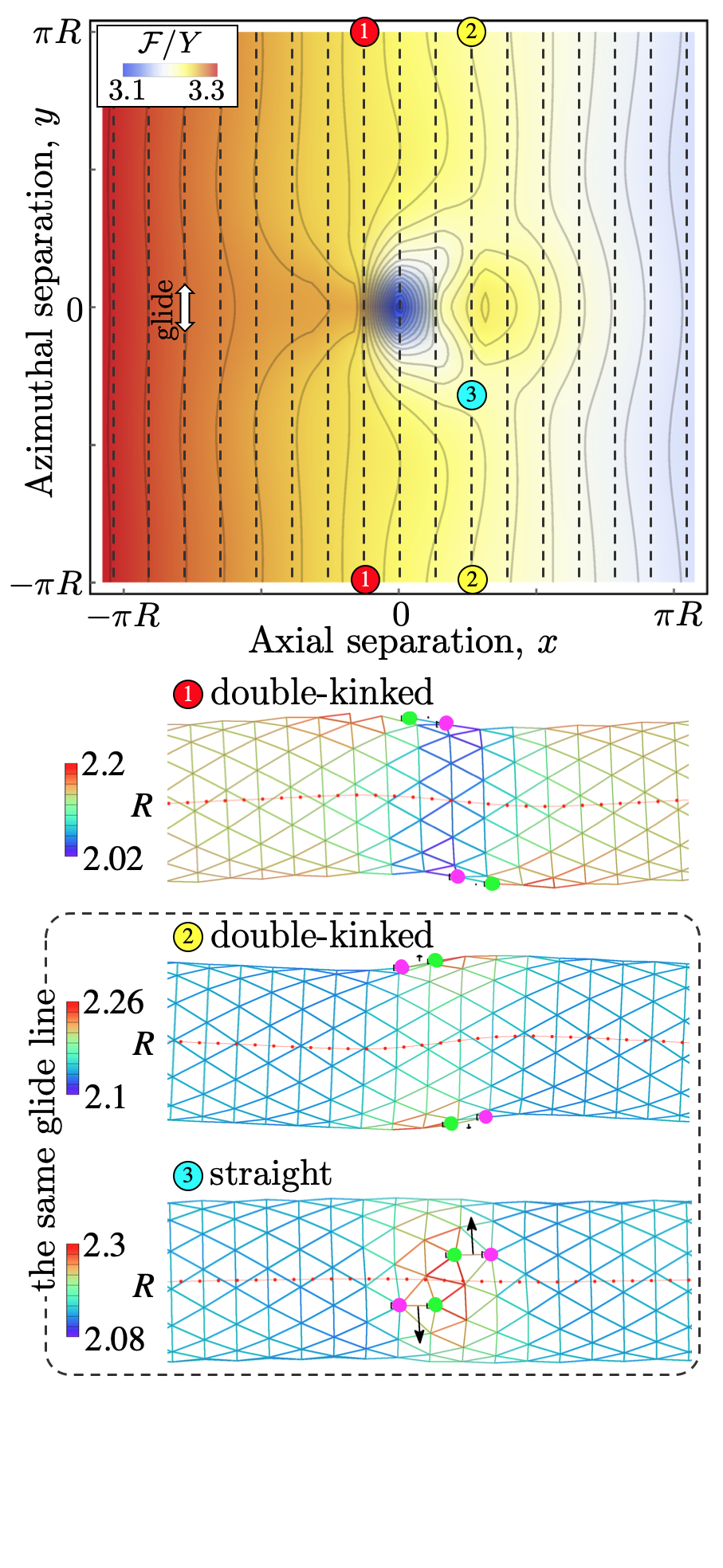}
			&\includegraphics[width=0.25\textwidth]{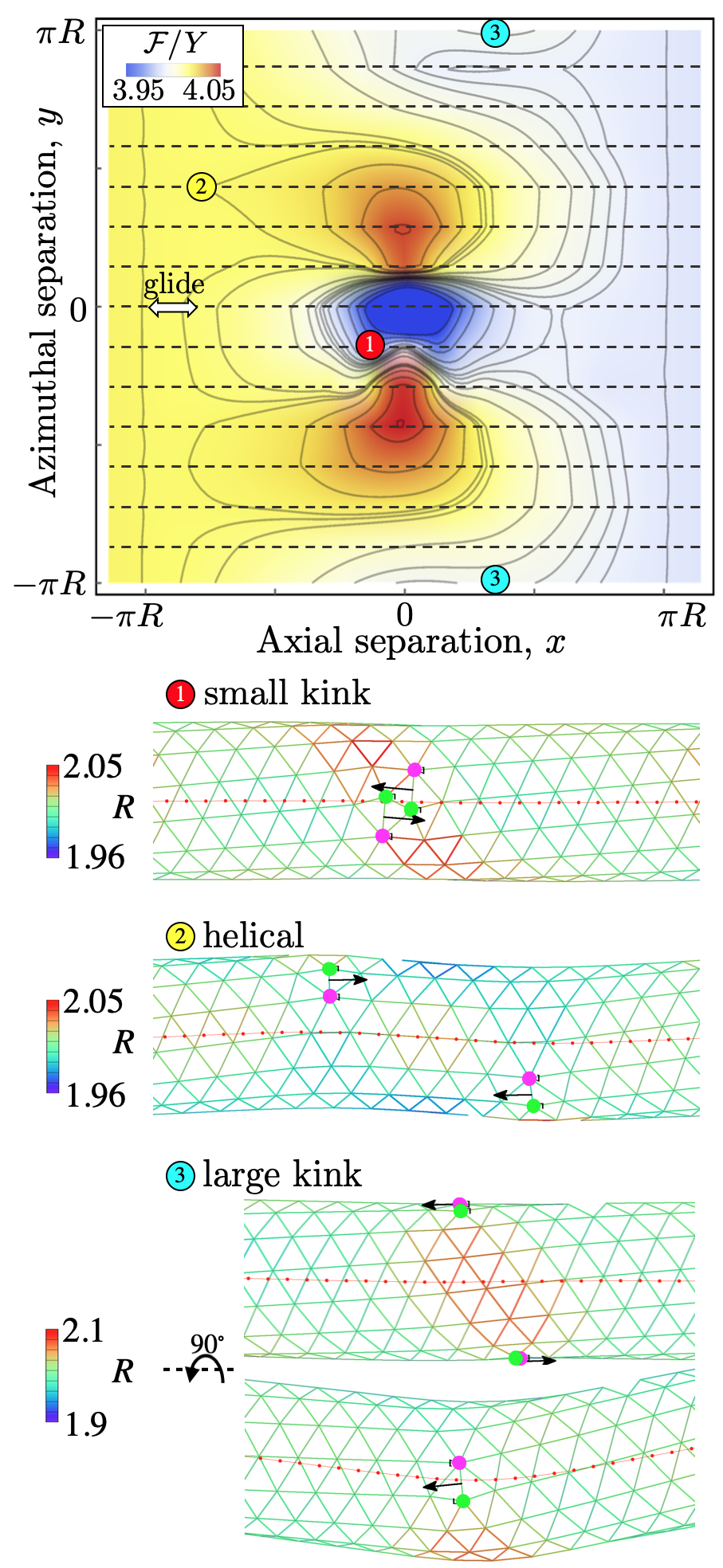}
			\end{tabular}
		\caption{Two interacting dislocation defects in a flexible tubular crystal ($\tilde{\kappa}=0.1$) with antiparallel Burgers vectors (black arrows) gliding (a) circumferentially  at $\theta=\pi/2$ in a zigzag lattice $(m,n)=(13,13)$ and (b) along the tube axis at $\theta=0$ in an armchair lattice $(m,n)=(14,7)$. The dashed lines in the energy landscapes (upper panels) correspond to the glide parastichies, whereas solid lines show contours of the energy landscape.  Three-dimensional conformations (lower panels), colored by a local measure of tube radius in units of the lattice spacing $a$, depict typical stable and metastable states marked in the energy plots.}
		\label{Fig:InterPair}
	\end{figure}

    As a first test case for dislocation pair-interactions in the freestanding tubular crystal,  we consider an $(m,n)=(13,13)$ tube with reduced bending rigidity $\tilde \kappa = 0.1$ and with a dislocation pair described by Burgers vector orientations $(\theta, \theta')= \pm \pi /2$.  The chosen dislocations are restricted to glide circumferentially along closed paths at constant $x$.     In Fig.~\ref{Fig:InterPair}a, we plot the numerically calculated energy landscape as a function of axial and azimuthal separations $x$ and $y$ between the defects, with $x$ held fixed in any one realization while glide alters $y$ (dashed lines). It is immediately apparent from Fig.~\ref{Fig:InterPair}a that the stable $y$-separations depend strongly on $x$, in a manner that lacks the $x\rightarrow -x$ symmetry of Eq.~\ref{Fan}.  An energy gradient from positive to negative $x$ arises from the bending energy's preference for dislocation motions that increase the tube radius \cite{BellerPRE16}; however, this gradient has no effect in our chosen example because $x$ is fixed. 
        
    Instead, we draw attention to the locations of energy minima along the glide lines of constant $x$. According to Eq.~\ref{Fan}, the fixed-cylinder case has no metastable states; there are either two equal minima, symmetrically placed around $y=0$, for $|x|<\pi R/2$, or one minimum at maximal azimuthal separation $y=\pi R$ for $|x|> \pi R/2$ (Fig.~9a in Appendix D). For the freestanding tube, we find a strikingly different energy landscape. At small $|x|$, the absolutely stable state is at $y=\pi R$, a configuration that the tube accommodates by taking on a double-kinked shape with oppositely oriented kinks at each dislocation (states 1 and 2 in Fig.~\ref{Fig:InterPair}a).  Another pair of local minima, with smaller $|y|$ symmetrically placed about $y=0$, are  metastable. In these configurations, disclinations of like sign are close to each other (state 3 in Fig.~\ref{Fig:InterPair}a), generating similar local surface deformations whose overlap is costly; meanwhile, the  tube axis remains approximately straight. 
    
	Notably, the metastable states at small $|y|$ are not symmetric about $x=0$. For $x<0$, where the negative disclinations are nearer to each other (i.e.\ the positive disclinations are on the outside), the region between the dislocations has  smaller radius and thus higher bending energy in  the vicinity of the defects, which effectively makes the dislocations more attractive, whereas for $x>0$ the bending energy pushes the dislocations apart in order to enlarge the tube's wider central region.
    
    {At larger $|x|$ (where $y=\pi R$ is stable in the rigid-cylinder case), here the absolutely stable state switches from $y=\pi R$ to the small-$|y|$ pair of states. A contributing factor in this swap is the decreased repulsion between dislocations, owing to their  overlapping surface deformations,  as their $|x|$ separation increases. Gradually, on further increasing $|x|$,  the $|y|$-locations of the absolutely stable states move out toward $|y|=\pi R$, eventually approaching the prediction of Eq.~\ref{Fan}. The special case of $x=0$, for which dislocations could pair-annihilate at $y=0$, unsurprisingly has a global minimum at $y=0$;  but, more interestingly, there still exists a metastable state at $y=\pi R$ with a kinked tube axis.
    Overall, our calculations show in this example that the freestanding tubular crystal has a richer landscape of effective defect interactions, with less symmetry and new metastable states, as compared with the crystal on a fixed cylinder.
    }
    
    For another illustrative example of metastability in dislocation interactions, we next examine a  dislocation pair gliding along the tube axis at $\theta=0$,  in an armchair lattice prescribed by $n=m/2=7$. Each dislocation now has fixed $y$ and variable $x$. The choice $n=m/2$ minimizes $R(m,n)$ for a given $m$ according to Eq.~\ref{Rmn}, which diminishes the decrease in bending energy associated with tube-widening glide moves. Therefore, we expect that for small $|x|$  the  stretching energy will dominate over the bending energy. In this regime, do defects in the freestanding tubular crystal act as they do on a fixed cylinder?  For the latter, the analytical solution (\ref{Fan}) predicts a pair of energy minima symmetric about $x=0$ for small fixed azimuthal separation $y$,  whereas for larger $y$ the dislocations repel indefinitely to large $|x|$ (Fig.~S2b). At $y=0$ the defects attract each other and annihilate at $x=0$. No metastable states are predicted at finite $x$. 

    Figure \ref{Fig:InterPair}b shows the energy landscape for the $\theta =0 $ dislocation pair in a freestanding tubular crystal. When both dislocations move along the same glide parastichy ($y=0$), they attract and annihilate, as in the fixed-cylinder case. However, whereas the energy landscape for the  fixed cylinder has a single global maximum at $(x,y) = (0,\pi R)$,  here we see a pair of maxima at $x=0$ situated near the $y=0$ minimum. This has a few consequences for stability when we fix $y$ in this landscape. The double minimum configuration exists only at $y=-a\sqrt{3}/2$, causing only small deflections from the initial shape (state 1 in Fig.~\ref{Fig:InterPair}b). Meanwhile, a new metastable state arises at intermediate azimuthal separations, as the inward-shifted energy maxima present a barrier to the bending energy's push in the $+x$-direction (state 2 in Fig.~\ref{Fig:InterPair}b). The tube shape is helically deformed in these states, a generic consequence of  kinks formed around defects at  azimuthal separations neither $0$ nor $\pi R$.  The energy barrier becomes smaller with increasing $y$, until eventually the barrier disappears and the defects are able to glide freely to the tube ends, pushed by the bending energy to increase $R$. However, when the defects are located on  opposite sides of the tube at $y=\pi R$, there exists another metastable state (state 3 in Fig.~\ref{Fig:InterPair}b), which creates a shape with a single large kink in the tube axis ($\approx 27^{\circ}$).
	
	The kink angle caused by a single dislocation can be predicted based on Burgers vector orientation $\theta$ and  tube radius $R$. The lattice contracts by one lattice spacing in the radial direction at $\theta=\pi/2$ with no change in chirality, and it shrinks by $a$ in the longitudinal direction at $\theta=0$. Since the longitudinal component of contraction, $a \cos{\theta}$, mainly contributes to the tube axis reorientation, a single dislocation causes a kink angle $\chi\approx \arcsin(a \cos{\theta}/(2R))$, which agrees well with the simulation results. 
    
    Extending this calculation to pairs of dislocations, we first note that two defects with parallel Burgers vectors on opposite sides of the tube cancel each others' effects on the tube axis orientation, whereas antiparallel $\mathbf{b}$ doubles the net kink angle. Defects at general azimuthal separation reduce each others' contributions to the kink angle by a factor $\cos(\beta)$, where $\beta$ is the angle between the Burgers vectors of the two dislocation defects in the cylindrical projection of the tube into a plane. Thus, our  prediction for the total kink angle reads $\chi_\mathrm{tot}\approx \arcsin(a \cos{\theta}/(2R))(1+\cos(y/R)\cos(\beta))$, which is in excellent agreement with the simulation results $\chi_\mathrm{tot}\approx 27.5^{\circ}$ at $\theta=0$, $R\approx2.1\,a$, $y=\pi R$, $\beta=\pi$ corresponding to state 3 depicted in Fig.~\ref{Fig:InterPair}b.  
	
    The two $(m,n)$ examples we have presented so far reveal metastable states under dislocation glide that are unique to the freestanding tubular crystal. These states produce (and are stabilized by) kink-bent, double-kinked, or helically kinked tube conformations in competition with straight or nearly straight conformations. 
	
\section{Control parameters for multistability}

    We now demonstrate that our principle of dislocation-mediated shape multistability is  versatile, as it can be controlled by a number of parameters. Some of these parameters could conceivably be changed dynamically during an experiment, such as through external forces or a temperature dependence of material constants. 

\subsection{Bending rigidity}

	\begin{figure}[t]
		\centering
		\setlength{\tabcolsep}{0.0em}
		\begin{tabular}{cc}
			(a)&(b)\\
			\includegraphics[width=0.24\textwidth]{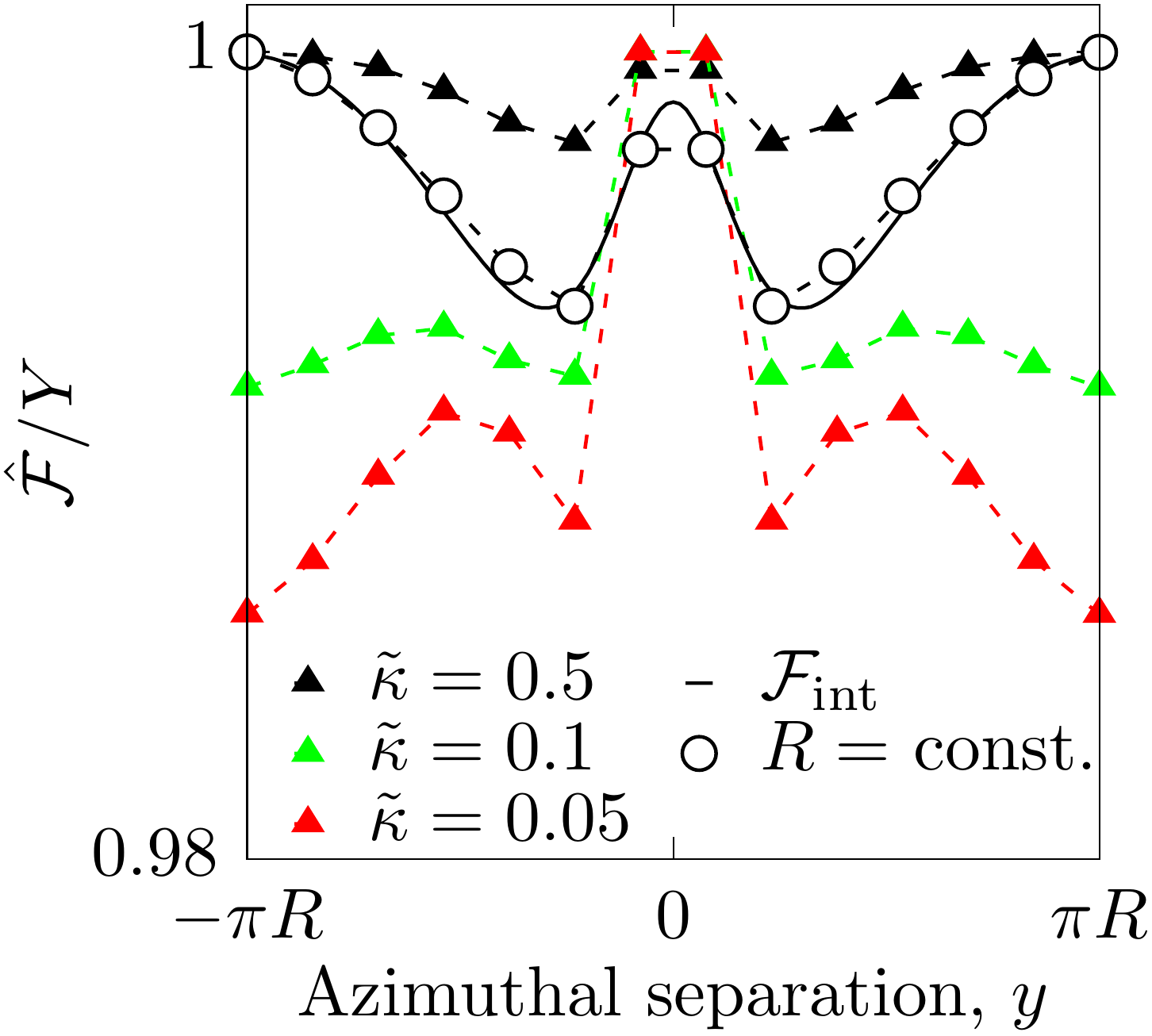} 
			&\includegraphics[width=0.235\textwidth]{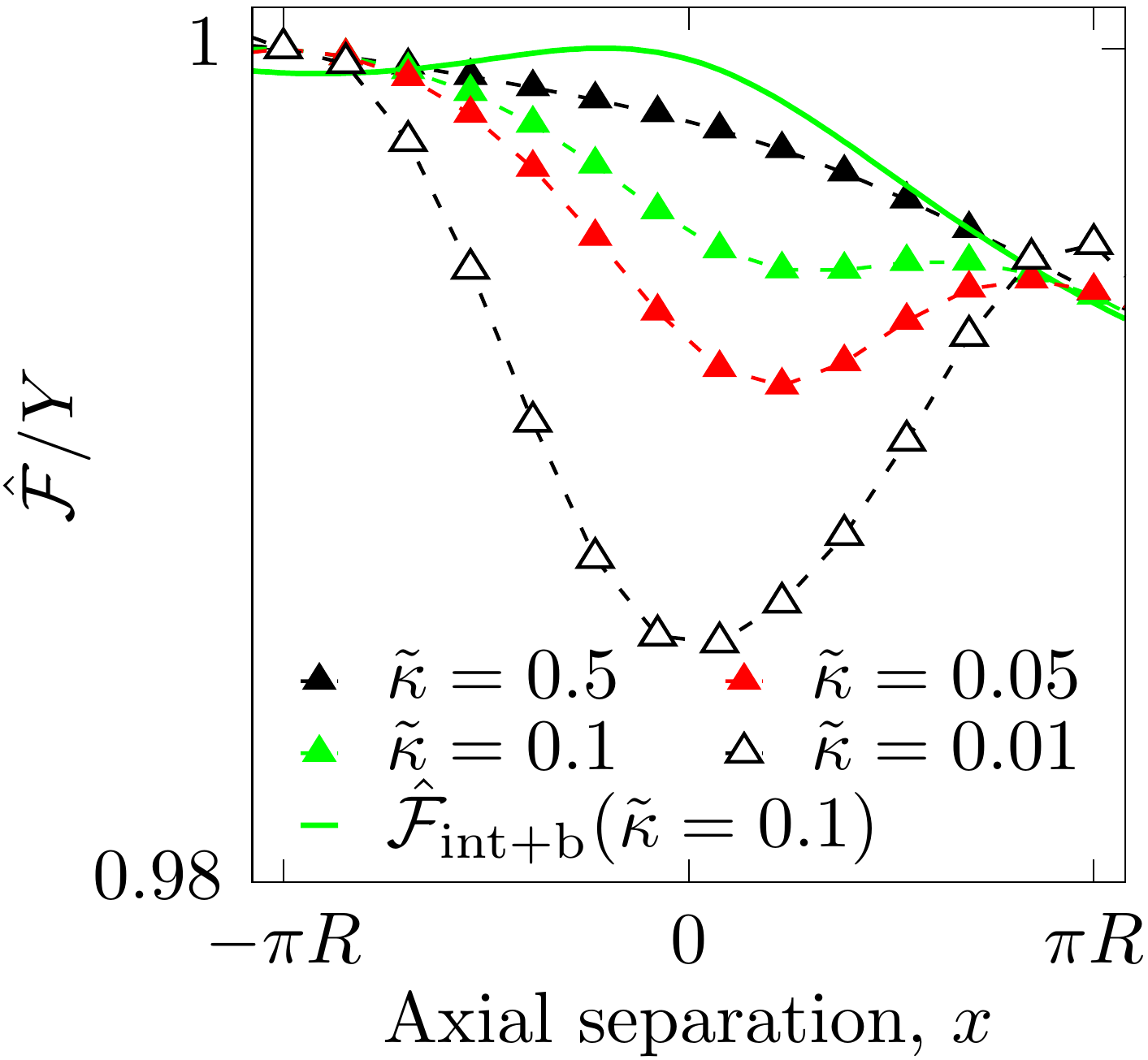}
		\end{tabular}
		\begin{tabular}{cc}	   
			(c)&(d) \\
		    \includegraphics[width=0.24\textwidth]{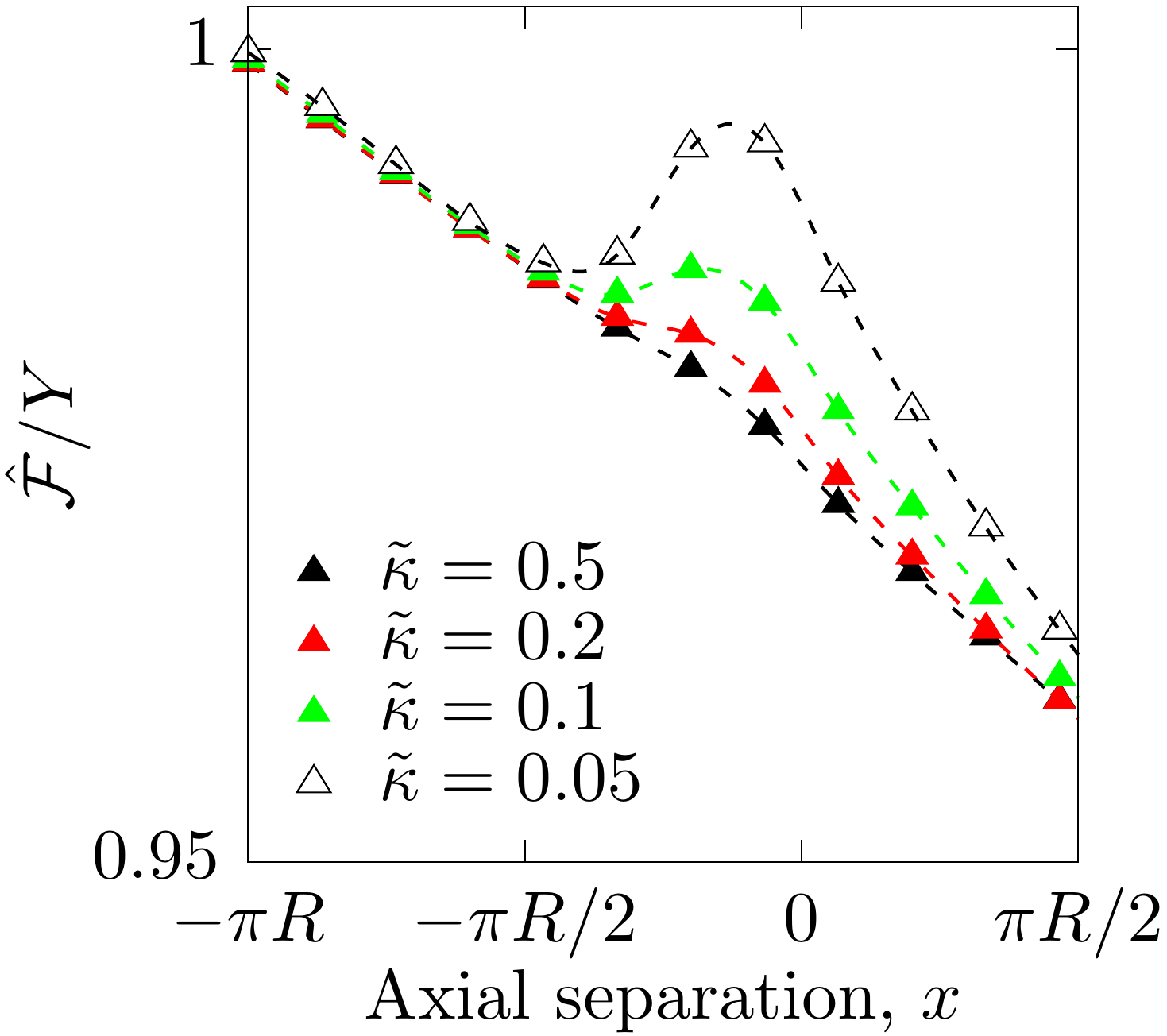}
			&\includegraphics[width=0.24\textwidth]{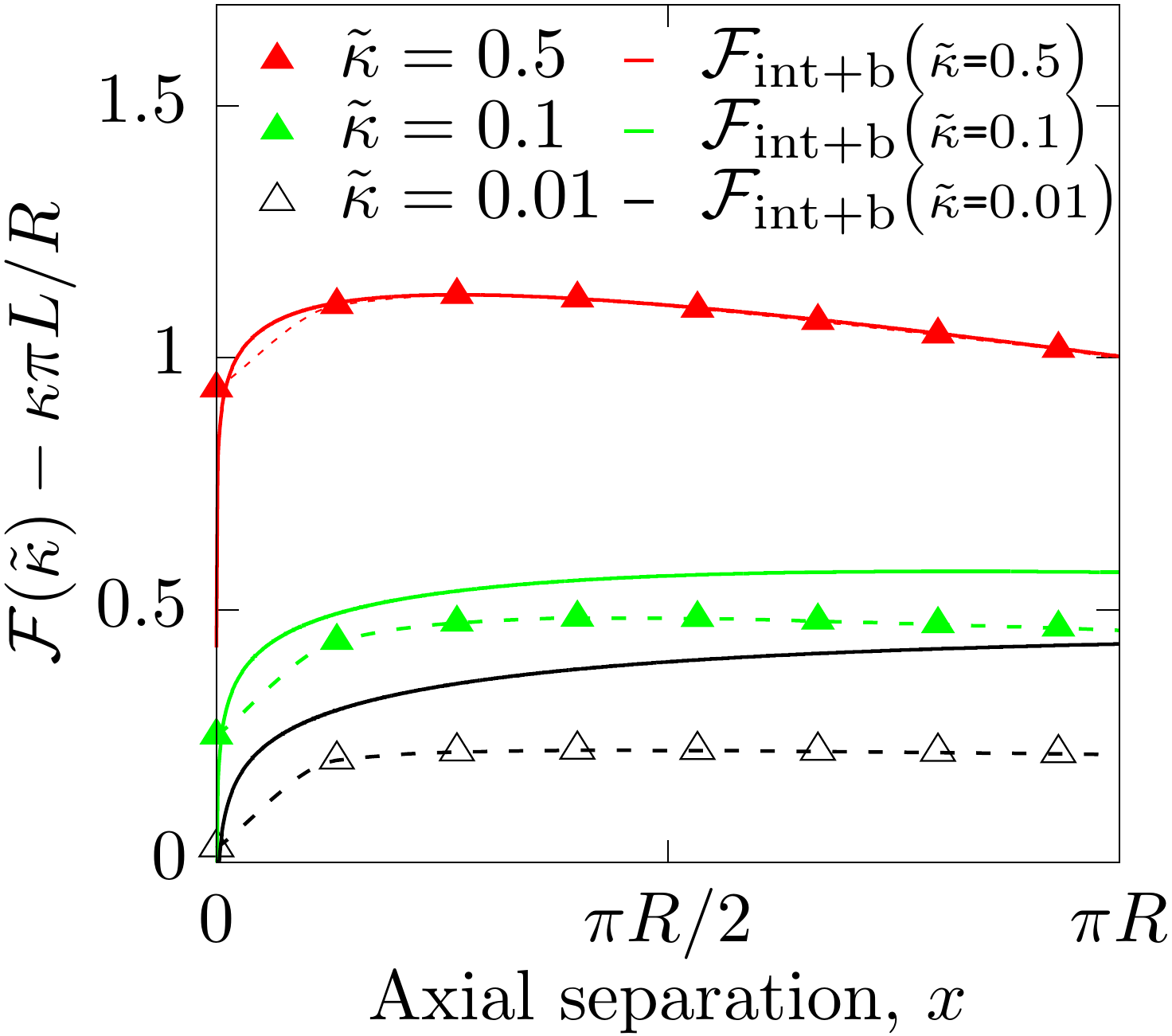}
		\end{tabular}
		\caption{
		Energy profiles for two interacting, gliding dislocations, at varying values of reduced bending rigidity $\tilde{\kappa}$, with (a) $\theta=\pi/2$ and $x=a\sqrt{3}$ in an $(m,n)=(13,13)$ tube;  (b) $\theta=0$ and $y=\pi R$ in an $(m,n)=(14,7)$ tube; and (c) $\theta=\pi/6$ and $c=a\sqrt{3}$ climb-steps separation in an $(m,n)=(13,13)$ tube.  The energy is normalized to a rescaled energy $\mathcal{\hat{F}}/Y$ in each plot such that the largest value is 1. (d) Dislocations moving along the same glide parastichy  at $\theta=\pi/6$ in the $(m,n)=(13,13)$ tube are attractive only at small separation and then become repulsive (triangles), whereas a na\"ive analytical approach (solid curves) predicts pair annihilation from any initial separation for small $\tilde{\kappa}$.    
		}
		\label{Fig:rigidity}
	\end{figure}

    First we examine how the energy barrier and the difference in energy between stable configurations depend on reduced bending rigidity $\tilde \kappa$. Smaller $\tilde \kappa$ favors metastable states in two ways: by permitting localized tube deformations to accommodate nearby dislocations, and by reducing the tube-widening stress that pushes defects to infinite separation. Thus, as we show below, changing $\tilde \kappa$ \textit{in situ} can cause a snap-through transition between two (meta)stable tube shapes or destabilize a formerly metastable state. 
    
    Returning to the  $(m,n)=(13,13)$ tube, for two dislocations with azimuthally-oriented Burgers vectors  ($\theta,\theta'=\pm \pi/2$) we test the multistability at $x=a\sqrt{3}$ noted above (states 2 and 3 in Fig.~\ref{Fig:InterPair}a) under changing $\tilde \kappa$. As shown in Fig.~\ref{Fig:rigidity}a,  the configuration at separation $y=\pi R$,  which is absolutely stable in a soft lattice  at small $\tilde{\kappa}$ (red triangles in Fig.~\ref{Fig:rigidity}a), becomes metastable at intermediate $\tilde \kappa$  and then unstable  when $\tilde{\kappa}>0.25$, in favor of the competing stable configuration at small $|y|$. For comparison, we also show the corresponding calculation for the fixed-cylinder case: By constraining all particles to lie in a cylindrical surface, we obtain excellent agreement with the analytical prediction of Fig.~\ref{Fan} (open circles, black solid line in Fig.~\ref{Fig:rigidity}a). In this example, the fixed-cylinder crystal follows the trend observed with increasing $\tilde \kappa$ in the freestanding case, as $y= \pi R$ is highly unstable in favor of small $|y|$. 
    
    A similarly significant $\tilde \kappa$-dependence can be seen in the highly bent shape observed in the $(m,n)=(14,7)$ tube for $\theta=0$ and fixed $y=\pi R$, which as noted above has a metastable configuration at small, positive $x$ (state 3 in Fig.~\ref{Fig:InterPair}b). We show in Fig.~\ref{Fig:rigidity}b that this state is very stable for small $\tilde \kappa$, then weakly metastable for intermediate $\tilde \kappa$, and finally  unstable at large $\tilde \kappa$ requiring the defects to glide apart indefinitely. The fixed-cylinder analytical prediction predicts a metastability  near $x\approx -\pi R$, not seen in our simulations of freestanding tubes,  due to an energy barrier centered at $x=0$. Interestingly, the predicted $\mathcal{F}_\mathrm{int}$ from Eq.~\ref{Fan} retains starkly different stable states from the computed results even when we add a na\"ive version of the bending energy per unit length, $\mathcal{F}_b /L \rightarrow \pi \kappa /R_0$, using Eq.~\ref{Rmn} for $R_0$ with $(m,n)$ changing abruptly at the $x$-values of dislocations. The resulting total free energy prediction (green solid curve in Fig.~\ref{Fig:rigidity}b), is so different from the computed free energy at the same $\tilde \kappa$ that the positions of maximum and local minimum are almost swapped. 
 
	So far we have examined only azimuthal and axial glide trajectories, but shape multistability arises also in the more generic case of glide along helical parastichies. As an example, we examine the case of $\theta=\pi/6$ in the $(m,n)=(13,13)$ tube, with constant climb separation $c=2$. We find a local energy minimum at $x\approx -\pi R$ (Fig.~\ref{Fig:rigidity}c), in which the tube morphs into a helical shape, whereas  a straight shape is recovered if the defects surmount an energy barrier at small $|x|$ to glide indefinitely toward $x\rightarrow + \infty $.  This energy barrier  increases with decreasing $\tilde{\kappa}$, and similarly to the $\theta=0$ case, the metastable state can be attained only if the initial separation $x < - \pi R/2$. 
    
    If we instead choose $c=0$, allowing dislocation pair-creation or pair-annihilation,  there is an energy barrier at small $|x|$  that divides a short-ranged attractive region, dominated by the stretching energy, from a long-ranged repulsive region, dominated by the tube-widening stress from the bending energy (red triangles in Fig.~\ref{Fig:rigidity}d). Indeed, our na\"ive bending energy term summed with the analytical stretching energy of  Eq.~\ref{Fan} fits the data very well (red line in Fig.~\ref{Fig:rigidity}d). However, this agreement breaks down qualitatively at smaller $\tilde \kappa$, where the analytical approach predicts attraction at all $x$ due to the weakened repulsive influence of the bending energy (green, black solid lines in Fig.~\ref{Fig:rigidity}d). Instead, the computed results for the freestanding tube show that the dislocations become weakly repulsive for separations beyond just a few glide steps (green, black triangles in Fig.~\ref{Fig:rigidity}d). Evidently, the freedom to deform from the cylindrical reference state allows the freestanding tubular crystal to partially screen the interactions predicted by Eq.~\ref{Fan}.

\subsection{External bending stress}

	\begin{figure}[t]
		\centering
		    (a)\\
			\includegraphics[width=0.48\textwidth]{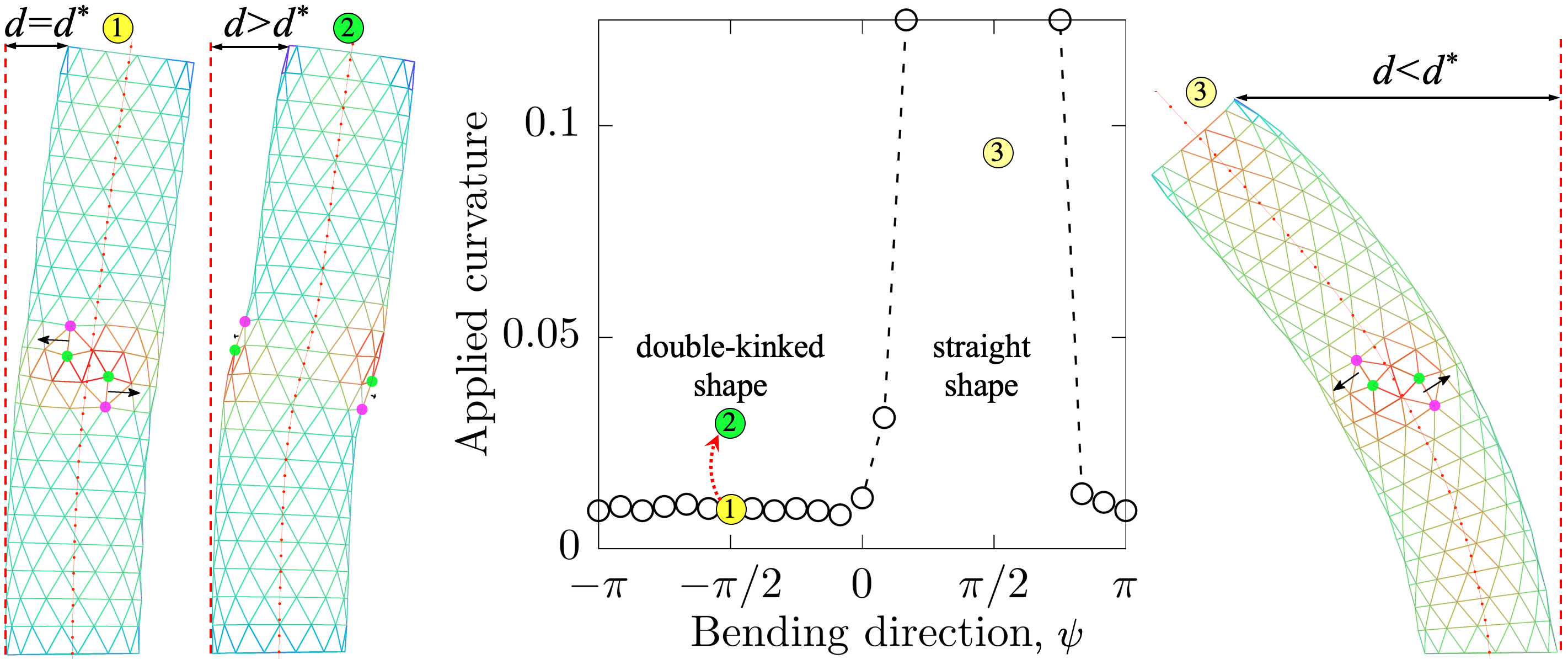}\\
			\begin{tabular}{cc}
			(b)&(c)\\
			\includegraphics[width=0.225\textwidth]{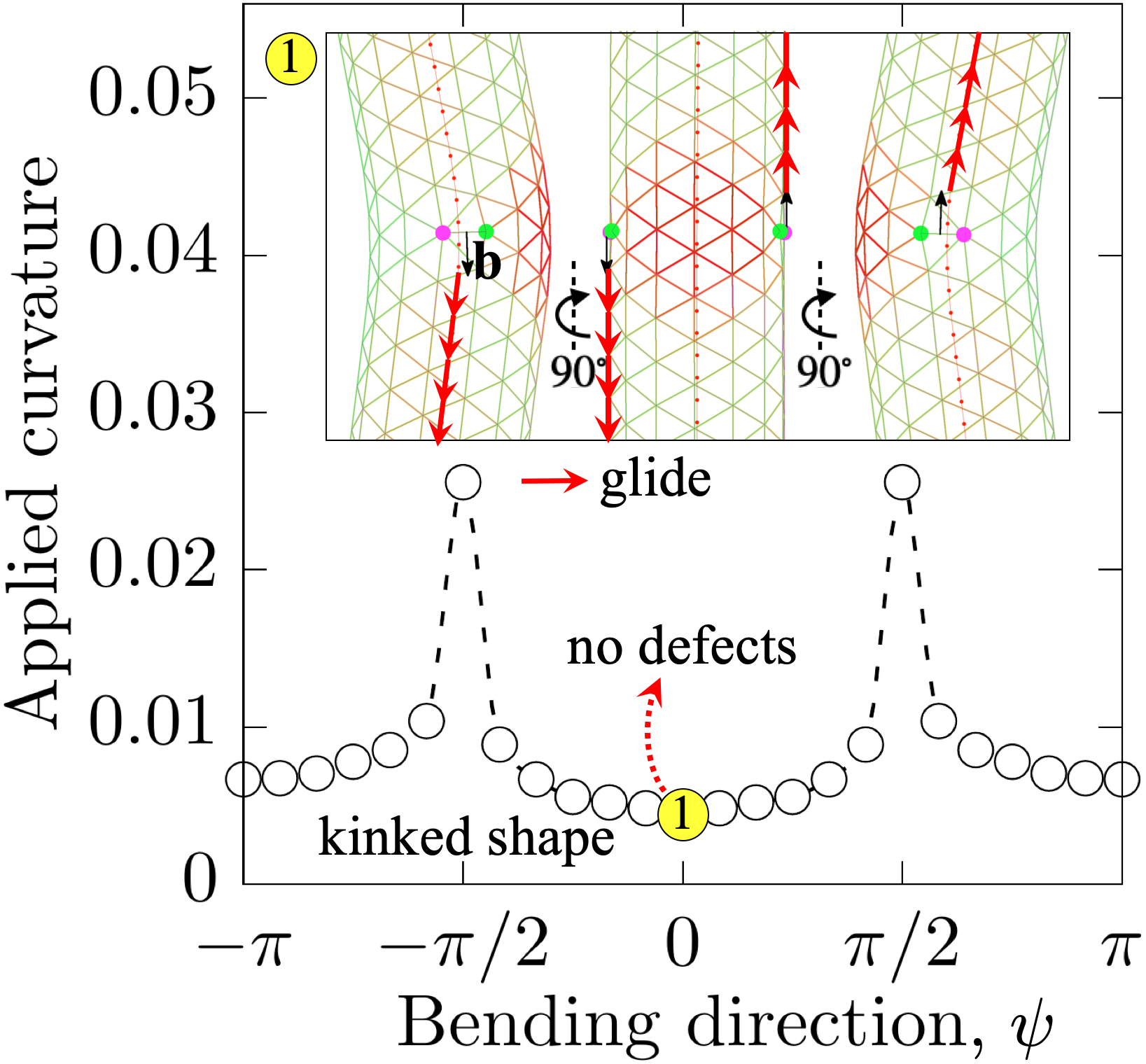} 
			&\includegraphics[width=0.225\textwidth]{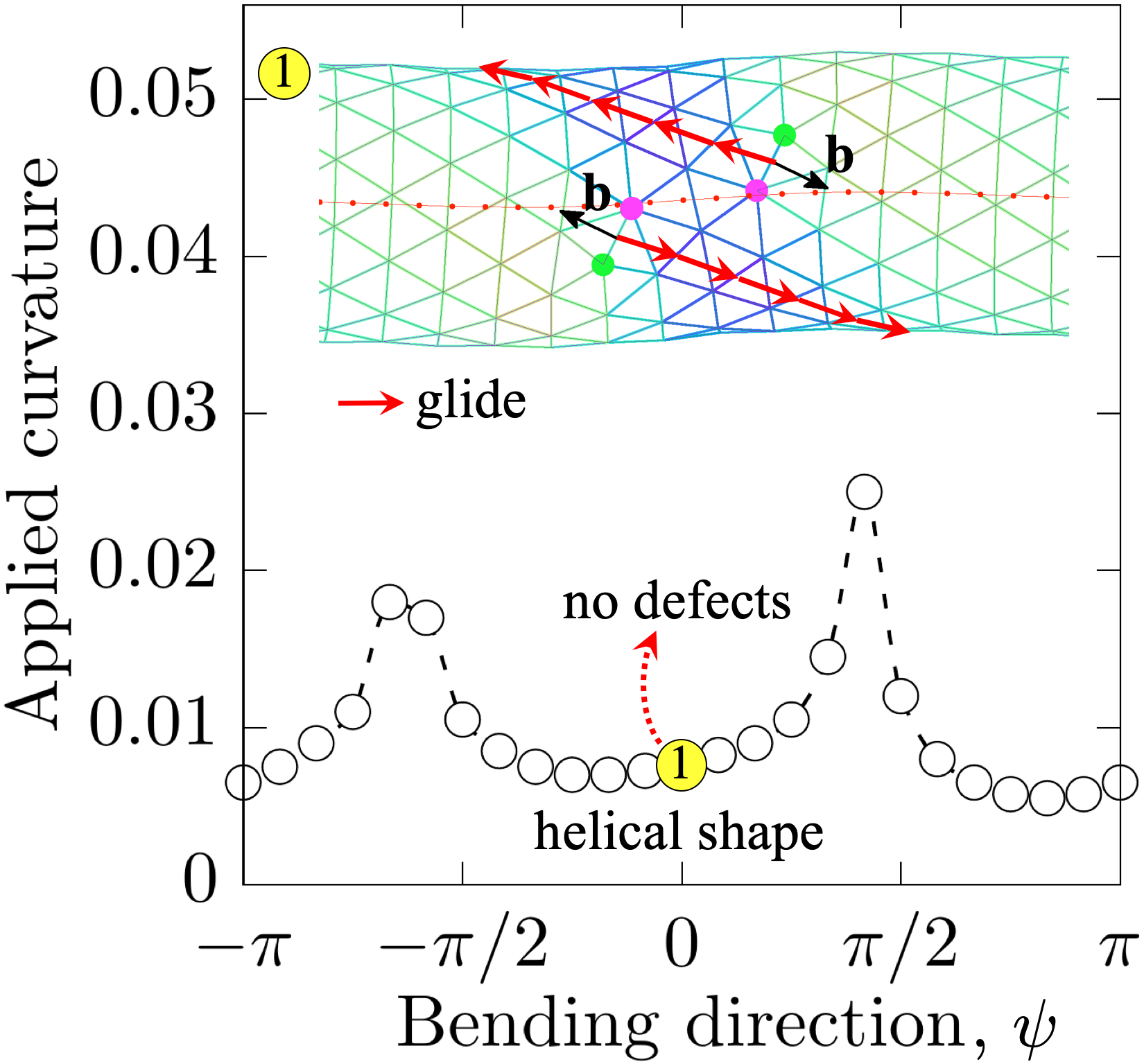}
		\end{tabular}
		\caption{ 
		Shape transition between stable states depending  on the orientation $\psi$ and magnitude (transverse displacement $d$) of applied bending strain at $\tilde{\kappa}=0.1$. Here $\psi=0$ corresponds to the average azimuthal position of the two dislocations being located on the compressed side of tube. The open circles (with dashed lines as guides to the eye) correspond to the critical displacement $d^*$ at which the shape change occurs. The red arrows in the insets depict the glide parastichies.
		(a) $\theta=\pi/2$ with $(m,n)=(13,13)$, (b) $\theta=0$ with $(m,n)=(14,7)$, and (c) $\theta=\pi/6$ with $(m,n)=(13,13)$.}
		\label{Fig:ApplBending}
	\end{figure}

    Externally imposed stresses present another set of routes to switching a freestanding tubular crystal between different (meta)stable states.  Here we focus on external bending stress as a means of overcoming the energy barriers described above.  The timescale for these switches is that of several glide steps, which we assume to be rapid enough to be observable \cite{lipowsky2005direct}. We take as initial states  some of the stable states identified above. To apply bending stress, the edge at one end of the tube is held fixed (taken to be clamped) while the other end is displaced from its reference position by distance $d$, producing curvature along the tube of $\approx2d/L^2$.  

    We first choose an angle $\psi$ in the $YZ$ plane to give, along with the initial tube axis $X$, a plane of bending, such that compression is maximum along the direction picked out by $\psi$.  For convenience, we take $\psi =0$ to be the average initial azimuthal position of the two dislocations. Then we incrementally increase the imposed bending curvature from zero, checking the stability of both dislocations at each increment and performing glide steps whenever this lowers the total energy. For each bending direction $\psi$, we thus find the critical curvature at displacement $d^{\ast}$ that destabilizes the initial defect configuration.  

    A particularly interesting case in which to study imposed bending is that of $(m,n)=(13,13)$, $\theta = \pi /2$ and $c\neq 0 $, as the dislocation pair can neither reach zero separation nor glide apart beyond a certain maximum distance. We found above  that a pair of small-$|y|$ stable states, causing a nearly straight tube shape, are accompanied at smaller $\tilde \kappa$ by a stable state at $y=\pi R$, which gives the tube a double-kinked shape   (Fig.~\ref{Fig:rigidity}a). The straight and double-kinked shapes have approximately equal energy at $\tilde \kappa  = 0.1$.   Using this value of $\tilde \kappa$, in Fig.~\ref{Fig:ApplBending}a we examine a tube with two dislocations at a small climb separation $c=2$, $x=a\sqrt{3}$ and an initially small azimuthal separation corresponding to one of the stable straight shapes (state 1). We find that the applied curvature necessary to enable the dislocations to glide apart, and produce the double-kinked shape (state 2), is highly dependent on bending direction $\psi$; for a large $\psi$-interval, the straight shape remains stable even at large applied curvature (state 3). For other $\psi$-values, bending easily produces a double-kinked shape which, by virtue of its energetic metastability, persists when the applied stress is removed. It is possible to reverse this transition, recovering the straight shape, by imposing bend in the other direction. 
   
    Interesting similarities and differences in this direction-dependent response to bending are seen when we examine  the singly kinked shape at $(m,n)=(14,7)$, $\theta=0$, $x=0$, $y=c=\pi R$,  $\tilde \kappa=0.1$ (state 1 in Fig.~\ref{Fig:ApplBending}b), which already resembles a bent configuration even without external forces. In this case, for all values of $\psi$, a finite applied curvature is found at which the $x=0$ state becomes unstable; the dislocations then glide apart to the ends of the tube, leaving behind no defects and thus a straight tube shape. However, the  kinked shape is much more easily destabilized at $\psi = 0$ or $\pi$, when the dislocations are $90^\circ$ degrees away from the bending plane, than at $\psi =\pm \pi /2$, when the dislocations are in the bending plane. A similar transition to a defect-free state from a helical shape takes place in an $(m,n)=(13,13)$ tube with defects at $\theta=\pi/6$ and short separation (Fig.~\ref{Fig:ApplBending}c). Since the shape is helical, this case lacks the $\psi \rightarrow -\psi$ symmetry seen in Fig.~\ref{Fig:ApplBending}b. Taken together, the results in Fig.~\ref{Fig:ApplBending} suggest that the $\psi$-dependence of $d^*$ changes smoothly with $\theta$, such that the sharp peak at $\psi = \pi/2$ remains while the peak at $\psi =-\pi/2$ diminishes and finally disappears as $\theta$ increases from zero to $\pi/2$. 

\subsection{Effect of spontaneous curvature}

    So far, we have assumed that the crystal has no spontaneous curvature, meaning that its ground state would be planar if it were ``unzipped'' from its cylindrical topology. But a spontaneous mean curvature clearly aids the assembly of a tube from a sheet, so it is important to address how such a material property will affect defect-mediated shape multistability. A plausible mechanism for such spontaneous curvature in  microtubules is the presence of additional proteins that bind to adjacent protofilaments and change the angle at lattice contacts \cite{brouhard2014contribution}. This provides \textit{anisotropic} spontaneous curvature, with different preferred curvatures along principal directions. We similarly impose spontaneous curvature $1/R_0$ along one principal direction of the crystalline membrane, whereas the spontaneous curvature remains  zero along the other direction, so that the bending energy is minimized in the initial, cylindrical state of a pristine tubular crystal. Details of the numerical implementation are described in Appendix A. 
    
    In Fig.~\ref{Fig:SpontCurv} we illustrate with two examples that spontaneous curvature may change which tube shapes are stable and thus offers another potential control parameter. Without spontaneous curvature, two dislocations  in an $(m,n)=(13,13)$ tube, oriented at $\theta=\pi/2$ and separated by $x=-a\sqrt{3}$,  have two  states (mirror reflections of each other) at short azimuthal separation (Fig.~\ref{Fig:InterPair}a) that lead to a helical shape (state 1 in Fig.~\ref{Fig:SpontCurv}a). Adding spontaneous curvature $1/R_0$ causes a transition to an absolutely stable double-kinked shape with defect separation $y=\pi R$. Another effect of spontaneous curvature is shown in Fig.~\ref{Fig:SpontCurv}b for a tube with defects gliding along the tube axis at $\theta=0$, $y=\pi R$. The spontaneous curvature stabilizes the metastable state at small $x$ by increasing the energy barrier and reducing the bending energy loss due to glide (i.e.\ the slope of the energy's dependence on $x$ is decreased). However, the global minimum-energy state is still the defect-free configuration occurring at $|x|\rightarrow \infty$, which can be reached, for example, by applied bending.  Comparing Fig.~\ref{Fig:SpontCurv} to Fig.~\ref{Fig:rigidity}a,b, we see that the introduction of spontaneous curvature has a similar effect to decreasing $\tilde \kappa$, with regard to which states defect separations are stabilized or destabilized. This can be rationalized by recognizing that the reference state is now a minimum of the bending energy, so the energy penalty for deviations from the reference state are of higher order in small bending distortions compared with the case of zero spontaneous curvature, which is always far from its locally preferred minimum. 
    
    We anticipate that spontaneous curvature also changes effective defect interactions in a manner similar to varying $\tilde \kappa$ for other choices of Burgers vector orientation $\theta$ and phyllotactic indices $(m,n)$. We leave for future work the computation of full energy landscapes at a range of $H_0$ values, examining here only two special configurations.  Spontaneous curvature can also change locally and temporally in realistic scenarios, for example with temperature or with concentration of binding proteins \cite{alushin2010ndc80, wilson2008orientation}, providing an additional way to control defect interactions and stable shapes.

	\begin{figure}[t]
		\centering
			\begin{tabular}{cc}
			(a)&(b)\\
			\includegraphics[width=0.23\textwidth]{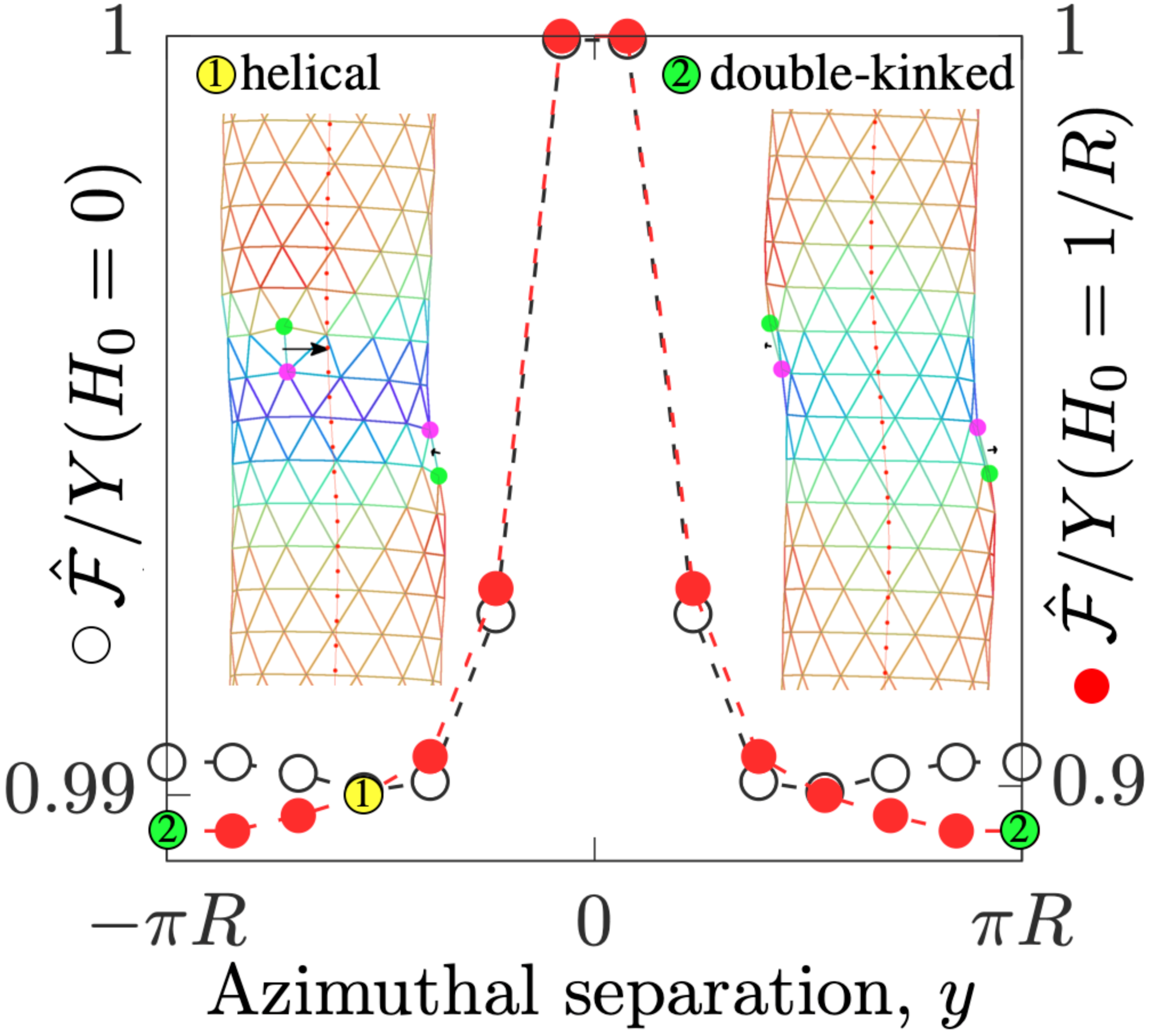} 
			&\includegraphics[width=0.23\textwidth]{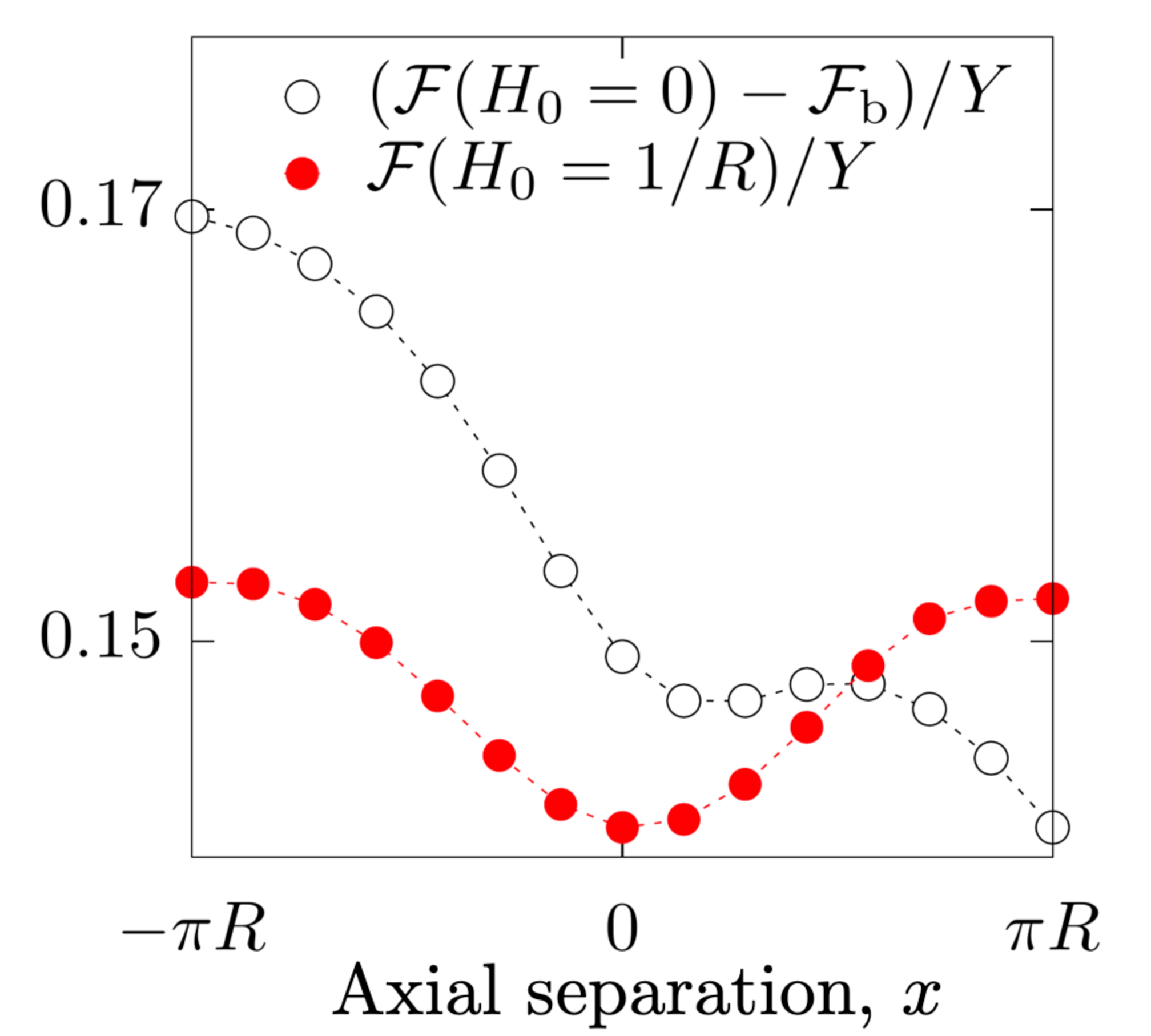}
		\end{tabular}
		\caption{Spontaneous curvature changes the optimal separation between two interacting dislocation defects, triggering a  transition  between stable helical and double-kinked shapes. Plots show the energy's dependence  on separation between dislocations for a tubular crystal of radius $R$ without (empty black circles) and with (filled red circles) spontaneous curvature $H_0=1/R$, with $\tilde{\kappa}=0.1$,  at (a) $\theta=\pi/2$, $x=-a\sqrt{3}$, $(m,n)=(13,13)$; and (b) $\theta=0$, $y=\pi R$, $(m,n)=(14,7)$. The bending energy of constant value $\mathcal{F}_b$ for a pristine tube  is subtracted in (b) from the total energy for the $H_0=0$ case in order to plot the considered cases on the same scale. }
		\label{Fig:SpontCurv}
	\end{figure}

\section{Designing target shapes via metastability and buckling}

	\begin{figure}[t]
		\centering
		(a)\\
			\includegraphics[width=0.48\textwidth]{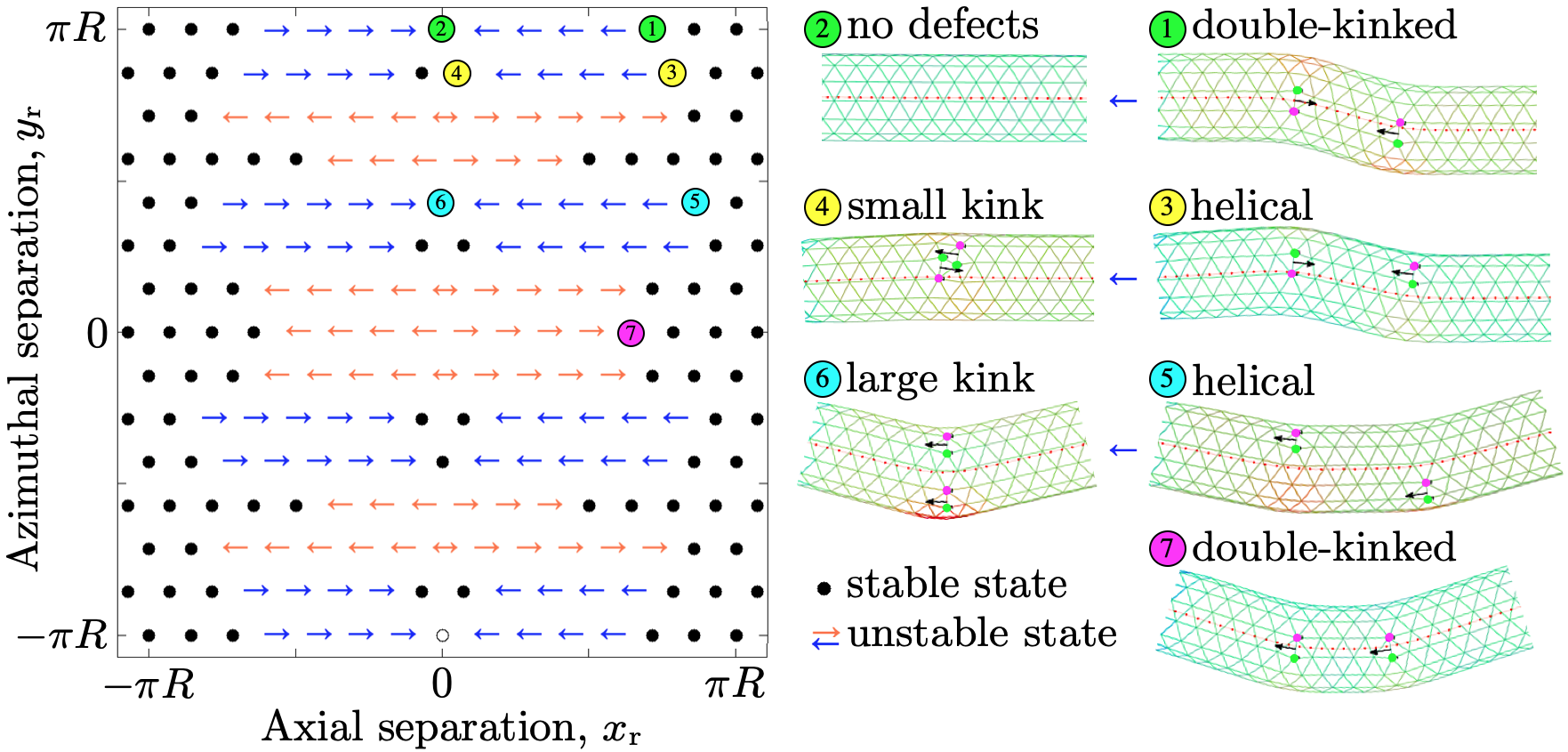} \\
			\setlength{\tabcolsep}{0.5em}
		\begin{tabular}{cc}
			(b)&(c)\\
			\includegraphics[width=0.23\textwidth]{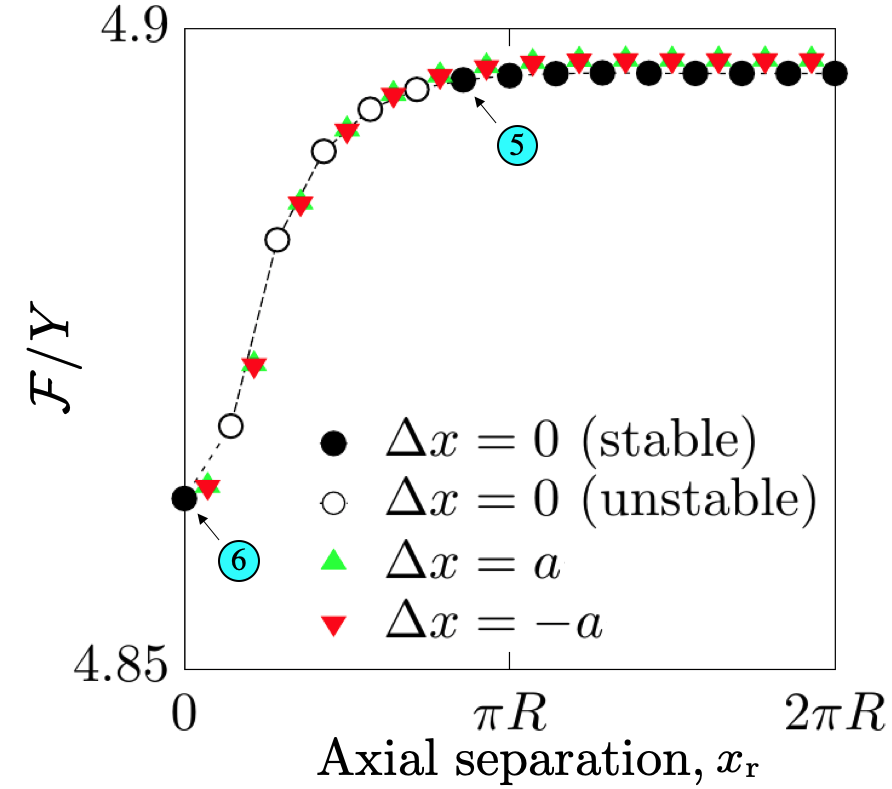} 
			&\includegraphics[width=0.23\textwidth]{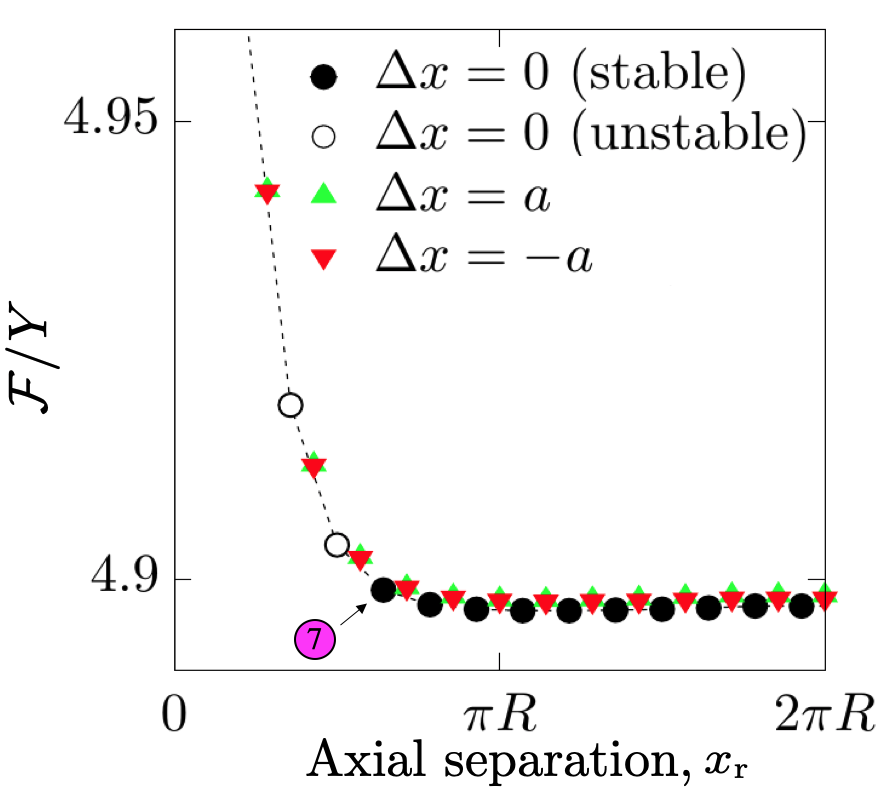}
		\end{tabular}	
			
			\caption{Stability of two pairs of dislocations gliding in an $(m,n)=(14,7)$ armchair tube with $\tilde{\kappa}=0.1$. (a) Stable (black dots) and unstable (arrows) relative displacements between the two dislocation pairs, with a locally stable $y=\pi R$, $x=0$ separation between the two dislocations within each pair. Collectively stable states exist at all possible relative azimuthal separation $y_\mathrm{r}$ at large relative separation $x_\mathrm{r}$ along the tube axis, but become unstable when separation $x_\mathrm{r}$ decreases, demonstrating attraction (blue arrows) to the absolutely stable state at small separation or  repulsion (orange arrows) to larger separation. Three-dimensional tube conformations are shown for several representative stable configurations on the right. (b) Energy profile for gliding at constant azimuthal separation $y_\mathrm{r}=3a\sqrt{3}/2$ between glide parastichies. At large $x_\mathrm{r}$, a glide step of either dislocation in the moving pair (red and green triangles) is unfavorable unless its partner glides in the same direction, giving a series of metastable states (solid black circles).  Decreasing the $x_\mathrm{r}$ separation eventually leads consecutive unstable states (open circles) as the two dislocation pairs attract toward the global minimum at $x_\mathrm{r}=0$, which gives the tube a large kink. Selected corresponding tube conformations  are depicted in (a). (c) Two dislocation pairs on the same glide parastichies at constant  $y_\mathrm{r}=0$. Pairs are stable (solid black circles) at large $x_\mathrm{r}$ and repulsive at small $x_\mathrm{r}$ separation (open circles). Shape at shortest separation is shown in (a). }
		\label{Fig:2pairsArmChair}
	\end{figure}
	\begin{figure}[t]
		\centering
		\includegraphics[width=0.5\textwidth]{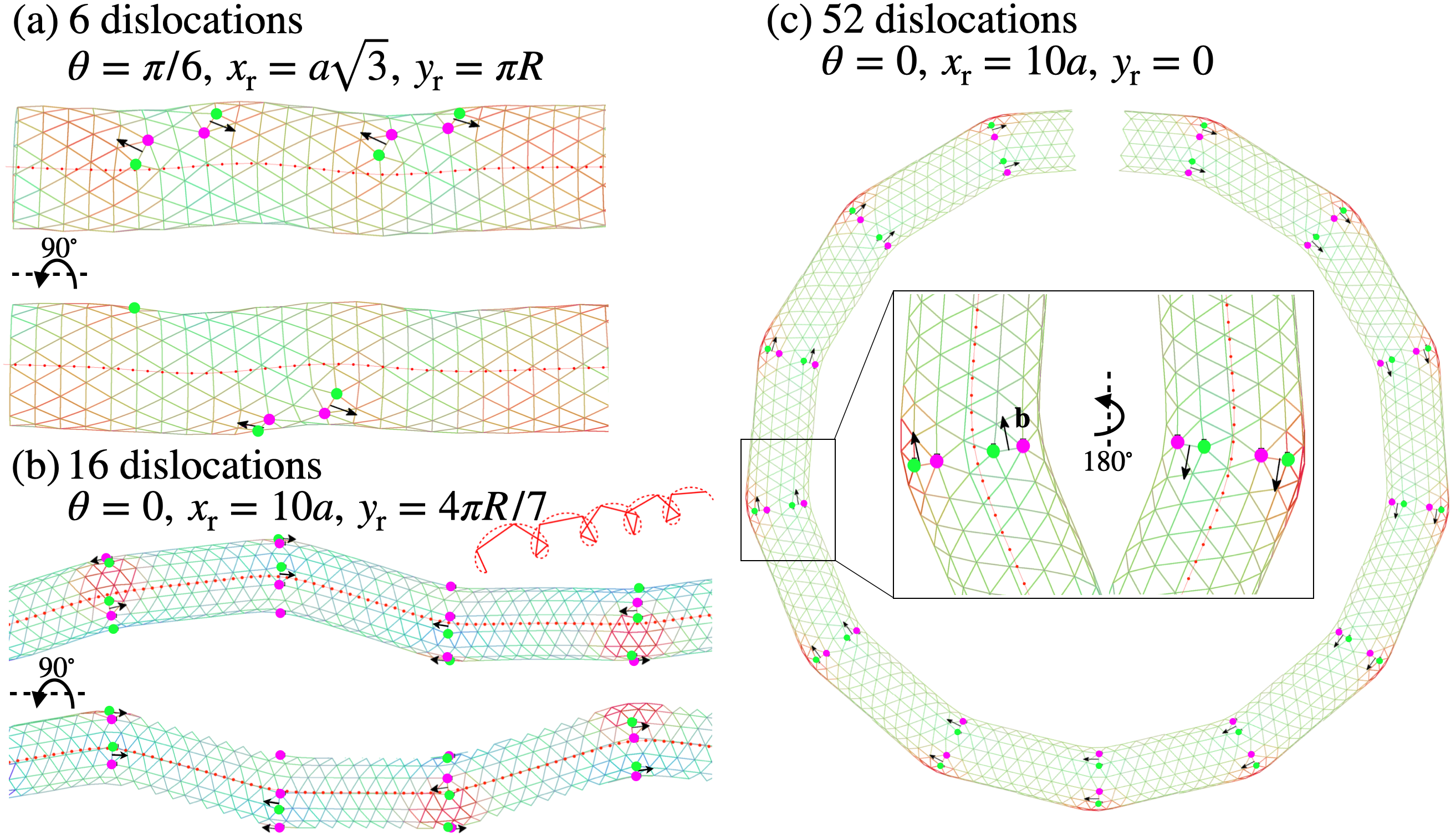} 
		\caption{Kinked tube shapes approximating continuously curved structures, 
		obtained by imposing a series of metastable dislocation pair groups at uniform relative separation ($x_\mathrm{r},y_\mathrm{r}$) and $\tilde{\kappa}=0.1$. (a,b) Tube develops a quasi-helical shape of different pitch and radius depending on the positions and orientations of defects. (c) Bent conformations obtained due to a series of repeating dislocation groups along the tube with no relative rotation about the tube axis ($y_\mathrm{r}=0$). The parastichy numbers are $(m,n)=(13,13)$ in (a), $(14,7)$ in (b,c).  }
		\label{Fig:Convoluted}
	\end{figure}
    We now build hierarchically upon our findings for dislocation pairs  to study interactions among multiple pairs of dislocations, as routes to targeting desired tube shapes over larger length scales. As we have shown, two dislocations with parallel gliding directions can be attracted to a number of possible stable configurations, depending on their climb separation and Burgers vector orientations, that significantly change the macroscopic shape of the tube. The main shape motifs we have found are kinked, double-kinked, helical, and (nearly) straight. We here take these shape elements as building blocks of programmable and switchable geometries at mechanical equilibrium.
	
    Because a generalization to interactions of arbitrarily many defects is impractical, we proceed by examining interactions of pairs of dislocation pairs, with each $\mathbf{b}$,$-\mathbf{b}$ dislocation pair in one of the stable states identified above for pair-interactions. The search space for stable configurations is then limited to just the relative $(x_\mathrm{r}$, $y_\mathrm{r})$ separation of dislocation pair 1 from dislocation pair 2. In Fig.~\ref{Fig:2pairsArmChair} we show a variety of stable states for dislocation pairs in an $(m,n)=(14,7)$ tube with $\theta = 0$  and constant separation $x=a\sqrt{3}/2$, $y=\pi R$ between the dislocations in each pair, each corresponding to state 3 in Fig.~\ref{Fig:InterPair}b. We initialize the two pairs of dislocation pairs at large $x_\mathrm{r}$ separation and then gradually probe the energy landscape by alternate gliding of the two dislocations in one of the pairs while the other pair is held fixed. By this means we identify the critical $x_\mathrm{r}$ separation at which the two dislocation pairs induce each others' relative glide motion, while each pair maintains the stable separation between its two dislocations. 
	
    Our simulations show that the dislocation pairs remain stable at any azimuthal separation when their $x_\mathrm{r}$-separation is large, $x_\mathrm{r} > \pi R$ (black dots in Fig.~\ref{Fig:2pairsArmChair}a). Since each defect pair causes a kink in the tube axis, the azimuthal separation defines the resulting tube shape. At $y_\mathrm{r}=\pi R$, for example, we obtain a double-kinked shape (state 1 in Fig.~\ref{Fig:2pairsArmChair}a) because the two pairs are oriented in opposite directions in the 3D embedding space, and thus the curvature at the two kinks has opposite sign. However, if $x_\mathrm{r} < a 2 \sqrt{3}$, the dislocation pairs become attractive and annihilate (state 2), which is possible because at $y_\mathrm{r}=\pi R$ they glide along the same pair of parastichies. The final state is then a straight, pristine tubular crystal. 
    
    Dislocations also move along the same glide parastichies when $y_\mathrm{r}=0$,  but now their Burgers vectors become parallel in the 3D space.  As a result, pairs of dislocation pairs are repulsive at small $x_\mathrm{r}$ and generate two identical kinks at larger separation (state 7). Intermediate azimuthal separations lead to stable helical shapes at large $x_\mathrm{r}$ (states 3 and 5). Helicity arises due to bending of the tube in a different direction at each kink. At smaller $x_\mathrm{r}$, pairs repel for some $y_\mathrm{r}$-values but attract for others. Repulsion stabilizes the helical shape, whereas attraction stabilizes a small-$x_\mathrm{r}$ shape  with a single kink (states 4 and 6). The latter case presents a pathway for a spontaneous shape transition from a helical to a kinked conformation, if a small perturbation were to destabilize the helical state.  
    
	In Fig.~\ref{Fig:2pairsArmChair}b,c we show energy profiles with respect to axial separation $x_{\mathrm{r}}$ at constant azimuthal separation $y_{\mathrm{r}}$. The stable states (filled black circles) correspond to the stable states in Fig.~\ref{Fig:2pairsArmChair}a, whereas unstable states (empty circles) are marked by arrows in Fig.~\ref{Fig:2pairsArmChair}a. For $x_\mathrm{r} \gtrsim \pi R$  we see an alternation of locally stable and unstable states revealing a slight energy barrier encountered in gliding one dislocation away from its partner by $\Delta x = \pm a$, before the partner takes a glide step in the same direction. The energy landscape of the pair of dislocation pairs thus resembles the periodic Peierls potential felt by an individual dislocation, under the assumption that the time between glide steps is always long compared to the timescale for elastic relaxation. Besides this small oscillation, the energy remains essentially constant at larger $x_{\mathrm{r}}$. 
    
    However, for smaller initial separation $|x_\mathrm{r}|<\pi R$, interactions between the two dislocation pairs overcome the energy barrier for glide steps, so that collective glide of the dislocation pair occurs spontaneously. For example, with a $y_\mathrm{r}=3a\sqrt{3}/2$ azimuthal separation between the two defect pairs, the two pairs approach until reaching the $x_\mathrm{r}=0$ state with a single kink  (state 6, Fig.~\ref{Fig:2pairsArmChair}b). On the other hand, when $y_\mathrm{r}=0$ the two dislocation pairs repel at short distance until reaching a stable configuration at $x_\mathrm{r}=5a$ with two kinks of the same orientation (state 7, Fig.~\ref{Fig:2pairsArmChair}c). 
    
    Having found stable configurations and corresponding tube shapes among pairs of dislocation pairs, we can now design more complicated geometries by adding dislocation-pair elements in a desired order. A series of stable dislocation pairs, arranged at constant $(x_\mathrm{r}, y_\mathrm{r})$ separation between consecutive pairs, creates repeating kinks along the tube. 
    Fig.~\ref{Fig:Convoluted}a depicts an $(m,n)=(13,13)$ tube with three metastable pairs with Burgers vector orientation $\theta=\pi/6$ relative to the tube axis. The first and third pairs  are located on one side of the tube whereas the middle pair is imposed on the opposite side  ($y_\mathrm{r}=\pi R$), and successive pairs have constant separation along the tube axis ($x_\mathrm{r}=a\sqrt{3}$). The combined effect of the three dislocation pairs leads to a tube shape that approximates a helical deformation.
	
    Larger deformations can be generated by imposing dislocations with Burgers vectors parallel to the  tube axis ($\theta=0$). In Fig.~\ref{Fig:Convoluted}b the $(m,n)=(14,7)$ tube has 16 dislocations that are localized in 4 equally separated groups along the tube ($x_\mathrm{r}=10a$),  each group consisting of two pairs and creating a large kink in the tube axis that corresponds to the absolutely stable state 6 in Fig.~\ref{Fig:2pairsArmChair}a. We impose dislocation groups at constant azimuthal separation ($y_\mathrm{r}=4\pi R/7$), which leads to rotations between successive kinks; as a result, the tube shape becomes approximately helical. The helical pitch and  radius can be controlled by changing the rotation angle and separation between dislocation groups,  and could even be varied along the length of the tube. The ratio between rotation angle and separation along the tube defines the torsion ($\approx y_r/x_r$), whereas the kink angle divided by separation approximates the curvature ($\approx \chi_\mathrm{tot}/x_r$) of the helical shape. If there is no azimuthal separation between consecutive defect groups, $y_\mathrm{r}=0$, then the tube bends in the same plane at each group (Fig.~\ref{Fig:Convoluted}c), and by imposing a series of metastable pairs we obtain tube shapes that approximate continuously curved planar geometries, such as a ring in the case of the torus-like structure of Fig.~\ref{Fig:Convoluted}d.

\section{Discussion}

    In this article, we have numerically demonstrated  an emergent phenomenon of dislocation-mediated shape multistability in freestanding, flexible crystals of tubular geometry. Our  simulations predict multiple metastable states restricting dislocation glide and causing macroscopic shape transitions, which make this system strikingly distinct from dislocations in a crystal attached to a rigid {cylindrical} substrate. We have explored this multistability by varying experimentally relevant design parameters for some chosen dislocation orientations and lattice helicity. Specifically, we found that changing the bending rigidity creates or eliminates certain metastable defect separations, and thus enables shape-morphing into nontrivial tube conformations, by altering the localization of surface deformations around the dislocations. Additionally, we showed that an external bending force allows dynamical control of dislocation motion between distinct stable configurations, enabling shape multistability even when material properties such as $\tilde \kappa$ are fixed.  We then demonstrated a new principle of shape programming by which imposed dislocation patterns generate shape-morphing into kinked shapes approximating targeted space curves. 

    The simulations presented in this work comprise just a few examples from a vast design space yet to be explored, opening multiple avenues for future investigations. Our approach can be generalized beyond  dislocation groups with collinear Burgers vectors to collections of  arbitrary  dislocation orientations, generating new degrees of freedom and crossing glide paths. We have examined only a few sets of phyllotactic indices, focusing on achiral states for simplicity, and leaving for future work a systematic survey of $(m,n)$ values---and coexisting sequences thereof---which may enable new tube shapes. We conjecture that chiral lattices will respond to bend in a manner that interpolates between our observations in the armchair and zigzag achiral tessellations, but that the response to imposed twist will change rapidly as $(m,n)$ are varied \cite{zhang2009dislocation}. Even more generally, dislocation-mediated shape multistability can be sought in other lattice types than the triangular lattice considered here, especially  honeycomb and rhombic tessellations as closer analogues to carbon nanotubes and microtobules. Unbound disclinations, while outside the   scope of this study because of their higher energy, are expected to further enrich the dislocation interaction landscape when they arise naturally on highly curved surfaces \cite{vitelli2006crystallography, bowick2009two}. Future exploration of these open questions will enable us to address the inverse problem: how to choose a prescribed defect pattern to obtain a desired, mechanically stable tube shape of arbitrary complexity.
    
    Molecular dynamics (MD) simulations of freestanding tubular crystals will be important in extending our findings from the elastic networks explored here to objects with excluded volume, such as colloidal particles. Such a framework would also enable investigation of the role of assembly kinetics in forming dislocations, such as during colloidal assembly or, potentially, the assembly of tubulin dimers into protofilaments and microtubules. Some effects that we have ignored for simplicity, such as thermal fluctuations and the Peierls potential, can be more naturally incorporated in the MD framework. Also, while we have assumed isotropic elasticity in this work, breaking this symmetry in the monomer interactions will likely lead to new phenomena relevant to protein assemblies such as microtubules. 

    The distinctions that we have emphasized between the fixed cylindrical crystal and the freestanding tubular crystal highlight the need for new analytical approaches, incorporating the interplay of effective defect interactions and the curvature of the crystalline surface as it dynamically adapts to the defects' presence. Our geometrical approach to calculating the tube axis kink angle is a step forward in this regard. However, computational approaches remain necessary for calculating mechanically stable tube shapes in the three-dimensional embedding space. 

    Inspired by the spherical crystal case, it is tempting to view the fixed-surface tubular crystal as a high-rigidity limit of the freestanding tubular crystal \cite{lidmar2003virus}. As we have explored and exploited in this work, that assumption is not always correct: the tube axis in the freestanding case can change its orientation at the dislocation site, by whatever angle best relieves the stress induced by the dislocation, with no cost in extra strain at large distance. Therefore, the effective interactions between defects must incorporate the changes they induce in the crystalline membrane's embedding in the 3D space. 

    The feedback that we observe between surface deformations and in-surface defect dynamics suggests connections to be explored between this work and other topics of current interest in soft matter physics, including: the stabilization of non-spherical droplet shapes in Pickering emulsions by packings of colloidal particles, often as crystals with many defects  \cite{XiePickering}; motile disclinations in active matter on flexible, curved surfaces \cite{keber2014topology,maroudas2020topologicalPublished}; and complex out-of-plane deformations in nematic elastomers with imprinted defects \cite{modes2011gaussian,zakharov2015reshaping,Gimenez_Pinto_2017}. We hope that our findings will spur experimental investigations of colloidal freestanding tubular crystals,  mesoscale analogues to carbon nanotubes and microtubules, as a versatile platform for programmable, reconfigurable microwires with switchable mechanical and photonic response properties \cite{Kim2012Fabrication,Solomon_2018}. 

\section*{Acknowledgments}
We thank Kinjal Dasbiswas and David R.\ Nelson for helpful comments. We gratefully acknowledge computing time on the Multi-Environment Computer for Exploration and Discovery (MERCED) cluster at UC Merced, which was funded by National Science Foundation Grant No. ACI-1429783.

\bibliography{arxiv}

\begin{thebibliography}{10}

\bibitem{thess1996crystalline}
A.~Thess, R.~Lee, P.~Nikolaev, H.~Dai, P.~Petit, J.~Robert, C.~Xu, Y.~H. Lee,
  S.~G. Kim, A.~G. Rinzler, {\em et~al.}, ``Crystalline ropes of metallic
  carbon nanotubes,'' {\em Science}, vol.~273, no.~5274, pp.~483--487, 1996.

\bibitem{klug1999tobacco}
A.~Klug, ``The tobacco mosaic virus particle: structure and assembly,'' {\em
  Philosophical Transactions of the Royal Society of London. Series B:
  Biological Sciences}, vol.~354, no.~1383, pp.~531--535, 1999.

\bibitem{nogales2001structural}
E.~Nogales, ``Structural insights into microtubule function,'' {\em Annual
  Review of Biophysics and Biomolecular Structure}, vol.~30, no.~1,
  pp.~397--420, 2001.

\bibitem{tymczenko2008colloidal}
M.~Tymczenko, L.~F. Marsal, T.~Trifonov, I.~Rodriguez, F.~Ramiro-Manzano,
  J.~Pallares, A.~Rodriguez, R.~Alcubilla, and F.~Meseguer, ``Colloidal crystal
  wires,'' {\em Advanced Materials}, vol.~20, no.~12, pp.~2315--2318, 2008.

\bibitem{yakobson1998mechanical}
B.~Yakobson, ``Mechanical relaxation and “intramolecular plasticity” in
  carbon nanotubes,'' {\em Applied Physics Letters}, vol.~72, no.~8,
  pp.~918--920, 1998.

\bibitem{adler1997history}
I.~Adler, D.~Barabe, and R.~V. Jean, ``A history of the study of phyllotaxis,''
  {\em Annals of Biology}, vol.~80, no.~3, pp.~231--244, 1997.

\bibitem{pennybacker2015phyllotaxis}
M.~F. Pennybacker, P.~D. Shipman, and A.~C. Newell, ``Phyllotaxis: Some
  progress, but a story far from over,'' {\em Physica D}, 2015.

\bibitem{li2005fabrication}
F.~Li, X.~Badel, J.~Linnros, and J.~B. Wiley, ``Fabrication of colloidal
  crystals with tubular-like packings,'' {\em Journal of the American Chemical
  Society}, vol.~127, no.~10, pp.~3268--3269, 2005.

\bibitem{sadoc2012phyllotaxis}
J.-F. Sadoc, N.~Rivier, and J.~Charvolin, ``Phyllotaxis: a non-conventional
  crystalline solution to packing efficiency in situations with radial
  symmetry,'' {\em Acta Crystallographica Section A: Foundations of
  Crystallography}, vol.~68, no.~4, pp.~470--483, 2012.

\bibitem{avan2020self}
Y.~Guo, B.~G.~P. van Ravensteijn, and W.~K. Kegel, ``Self-assembly of isotropic
  colloids into colloidal strings{,} bernal spiral-like{,} and tubular
  clusters,'' {\em Chem. Commun.}, vol.~56, pp.~6309--6312, 2020.

\bibitem{li2019self}
T.~Li, B.~Wang, J.~Ning, W.~Li, G.~Guo, D.~Han, B.~Xue, J.~Zou, G.~Wu, Y.~Yang,
  {\em et~al.}, ``Self-assembled nanoparticle supertubes as robust platform for
  revealing long-term, multiscale lithiation evolution,'' {\em Matter}, vol.~1,
  no.~4, pp.~976--987, 2019.

\bibitem{lohr2010helical}
M.~A. Lohr, A.~M. Alsayed, B.~G. Chen, Z.~Zhang, R.~D. Kamien, and A.~G. Yodh,
  ``Helical packings and phase transformations of soft spheres in cylinders,''
  {\em Physical Review E}, vol.~81, no.~4, p.~040401, 2010.

\bibitem{harris1980tubular}
W.~F. Harris and R.~O. Erickson, ``Tubular arrays of spheres: geometry,
  continuous and discontinuous contraction, and the role of moving
  dislocations,'' {\em Journal of Theoretical Biology}, vol.~83, no.~2,
  pp.~215--246, 1980.

\bibitem{chretien1992lattice}
D.~Chr{\'e}tien, F.~Metoz, F.~Verde, E.~Karsenti, and R.~H. Wade, ``Lattice
  defects in microtubules: protofilament numbers vary within individual
  microtubules,'' {\em Journal of Cell Biology}, vol.~117, no.~5,
  pp.~1031--1040, 1992.

\bibitem{mughal2011phyllotactic}
A.~Mughal, H.~K. Chan, and D.~Weaire, ``Phyllotactic description of hard sphere
  packing in cylindrical channels,'' {\em Physical Review Letters}, vol.~106,
  no.~11, p.~115704, 2011.

\bibitem{wood2013self}
D.~Wood, C.~Santangelo, and A.~Dinsmore, ``Self-assembly on a cylinder: A model
  system for understanding the constraint of commensurability,'' {\em Soft
  Matter}, vol.~9, no.~42, pp.~10016--10024, 2013.

\bibitem{mughal2014theory}
A.~Mughal and D.~Weaire, ``Theory of cylindrical dense packings of disks,''
  {\em Physical Review E}, vol.~89, no.~4, p.~042307, 2014.

\bibitem{fu2016hard}
L.~Fu, W.~Steinhardt, H.~Zhao, J.~E. Socolar, and P.~Charbonneau, ``Hard sphere
  packings within cylinders,'' {\em Soft Matter}, vol.~12, no.~9,
  pp.~2505--2514, 2016.

\bibitem{moon2004fabrication}
J.~H. Moon, S.~Kim, G.-R. Yi, Y.-H. Lee, and S.-M. Yang, ``Fabrication of
  ordered macroporous cylinders by colloidal templating in microcapillaries,''
  {\em Langmuir}, vol.~20, no.~5, pp.~2033--2035, 2004.

\bibitem{nisoli2009static}
C.~Nisoli, N.~M. Gabor, P.~E. Lammert, J.~Maynard, and V.~H. Crespi, ``Static
  and dynamical phyllotaxis in a magnetic cactus,'' {\em Physical Review
  Letters}, vol.~102, no.~18, p.~186103, 2009.

\bibitem{nisoli2010annealing}
C.~Nisoli, N.~M. Gabor, P.~E. Lammert, J.~Maynard, and V.~H. Crespi,
  ``Annealing a magnetic cactus into phyllotaxis,'' {\em Physical Review E},
  vol.~81, no.~4, p.~046107, 2010.

\bibitem{Amir13}
A.~Amir, J.~Paulose, and D.~R. Nelson, ``Theory of interacting dislocations on
  cylinders,'' {\em Phys. Rev. E}, vol.~87, p.~042314, Apr 2013.

\bibitem{BellerPRE16}
D.~A. Beller and D.~R. Nelson, ``Plastic deformation of tubular crystals by
  dislocation glide,'' {\em Phys. Rev. E}, vol.~94, p.~033004, Sep 2016.

\bibitem{dunlap1994constraints}
B.~I. Dunlap, ``Constraints on small graphitic helices,'' {\em Physical Review
  B}, vol.~50, no.~11, p.~8134, 1994.

\bibitem{bowick2009two}
M.~J. Bowick and L.~Giomi, ``Two-dimensional matter: order, curvature and
  defects,'' {\em Advances in Physics}, vol.~58, no.~5, pp.~449--563, 2009.

\bibitem{bausch2003grain}
A.~Bausch, M.~J. Bowick, A.~Cacciuto, A.~Dinsmore, M.~Hsu, D.~Nelson,
  M.~Nikolaides, A.~Travesset, and D.~Weitz, ``Grain boundary scars and
  spherical crystallography,'' {\em Science}, vol.~299, no.~5613,
  pp.~1716--1718, 2003.

\bibitem{lidmar2003virus}
J.~Lidmar, L.~Mirny, and D.~R. Nelson, ``Virus shapes and buckling transitions
  in spherical shells,'' {\em Physical Review E}, vol.~68, no.~5, p.~051910,
  2003.

\bibitem{vitelli2006crystallography}
V.~Vitelli, J.~B. Lucks, and D.~R. Nelson, ``Crystallography on curved
  surfaces,'' {\em Proceedings of the National Academy of Sciences}, vol.~103,
  no.~33, pp.~12323--12328, 2006.

\bibitem{ellis2018curvature}
P.~W. Ellis, D.~J. Pearce, Y.-W. Chang, G.~Goldsztein, L.~Giomi, and
  A.~Fernandez-Nieves, ``Curvature-induced defect unbinding and dynamics in
  active nematic toroids,'' {\em Nature Physics}, vol.~14, no.~1, pp.~85--90,
  2018.

\bibitem{seung1988defects}
H.~S. Seung and D.~R. Nelson, ``Defects in flexible membranes with crystalline
  order,'' {\em Physical Review A}, vol.~38, no.~2, p.~1005, 1988.

\bibitem{zhang2014defects}
T.~Zhang, X.~Li, and H.~Gao, ``Defects controlled wrinkling and topological
  design in graphene,'' {\em Journal of the Mechanics and Physics of Solids},
  vol.~67, pp.~2--13, 2014.

\bibitem{modes2011gaussian}
C.~Modes, K.~Bhattacharya, and M.~Warner, ``Gaussian curvature from flat
  elastica sheets,'' {\em Proceedings of the Royal Society A: Mathematical,
  Physical and Engineering Sciences}, vol.~467, no.~2128, pp.~1121--1140, 2011.

\bibitem{zakharov2015reshaping}
A.~Zakharov and L.~Pismen, ``Reshaping nemato-elastic sheets,'' {\em The
  European Physical Journal E}, vol.~38, no.~7, pp.~1--6, 2015.

\bibitem{giomi2008elastic}
L.~Giomi and M.~J. Bowick, ``Elastic theory of defects in toroidal crystals,''
  {\em The European Physical Journal E}, vol.~27, no.~3, pp.~275--296, 2008.

\bibitem{hirth1983theory}
J.~P. Hirth and J.~Lothe, {\em Theory of dislocations}.
\newblock Wiley, New York, 1982.

\bibitem{klein2007shaping}
Y.~Klein, E.~Efrati, and E.~Sharon, ``Shaping of elastic sheets by prescription
  of non-{E}uclidean metrics,'' {\em Science}, vol.~315, no.~5815,
  pp.~1116--1120, 2007.

\bibitem{gladman2016biomimetic}
A.~S. Gladman, E.~A. Matsumoto, R.~G. Nuzzo, L.~Mahadevan, and J.~A. Lewis,
  ``Biomimetic 4d printing,'' {\em Nature Materials}, vol.~15, no.~4,
  pp.~413--418, 2016.

\bibitem{aharoni2018universal}
H.~Aharoni, Y.~Xia, X.~Zhang, R.~D. Kamien, and S.~Yang, ``Universal inverse
  design of surfaces with thin nematic elastomer sheets,'' {\em Proceedings of
  the National Academy of Sciences}, vol.~115, no.~28, pp.~7206--7211, 2018.

\bibitem{brangwynne2007force}
C.~P. Brangwynne, F.~MacKintosh, and D.~A. Weitz, ``Force fluctuations and
  polymerization dynamics of intracellular microtubules,'' {\em Proceedings of
  the National Academy of Sciences}, vol.~104, no.~41, pp.~16128--16133, 2007.

\bibitem{janke2017causes}
C.~Janke and G.~Montagnac, ``Causes and consequences of microtubule
  acetylation,'' {\em Current Biology}, vol.~27, no.~23, pp.~R1287--R1292,
  2017.

\bibitem{hunyadi2007microtubule}
V.~Hunyadi, D.~Chr{\'e}tien, H.~Flyvbjerg, and I.~M. J{\'a}nosi, ``Why is the
  microtubule lattice helical?,'' {\em Biology of the Cell}, vol.~99, no.~2,
  pp.~117--128, 2007.

\bibitem{witten2007stress}
T.~A. Witten, ``Stress focusing in elastic sheets,'' {\em Reviews of Modern
  Physics}, vol.~79, no.~2, p.~643, 2007.

\bibitem{erickson1973tubular}
R.~O. Erickson, ``Tubular packing of spheres in biological fine structure,''
  {\em Science}, vol.~181, no.~4101, pp.~705--716, 1973.

\bibitem{landau_elasticity}
L.~D. Landau and E.~M. Lifshits, {\em Theory of Elasticity}.
\newblock Pergamon Press, New York, 1986.

\bibitem{schaedel2019lattice}
L.~Schaedel, S.~Triclin, D.~Chr{\'e}tien, A.~Abrieu, C.~Aumeier, J.~Gaillard,
  L.~Blanchoin, M.~Th{\'e}ry, and K.~John, ``Lattice defects induce microtubule
  self-renewal,'' {\em Nature Physics}, vol.~15, no.~8, pp.~830--838, 2019.

\bibitem{plummer2020buckling}
A.~Plummer and D.~R. Nelson, ``Buckling and metastability in membranes with
  dilation arrays,'' {\em Physical Review E}, vol.~102, no.~3, p.~033002, 2020.

\bibitem{lipowsky2005direct}
P.~Lipowsky, M.~J. Bowick, J.~H. Meinke, D.~R. Nelson, and A.~R. Bausch,
  ``Direct visualization of dislocation dynamics in grain-boundary scars,''
  {\em Nature Materials}, vol.~4, no.~5, pp.~407--411, 2005.

\bibitem{brouhard2014contribution}
G.~J. Brouhard and L.~M. Rice, ``The contribution of $\alpha$$\beta$-tubulin
  curvature to microtubule dynamics,'' {\em Journal of Cell Biology}, vol.~207,
  no.~3, pp.~323--334, 2014.

\bibitem{alushin2010ndc80}
G.~M. Alushin, V.~H. Ramey, S.~Pasqualato, D.~A. Ball, N.~Grigorieff,
  A.~Musacchio, and E.~Nogales, ``The ndc80 kinetochore complex forms
  oligomeric arrays along microtubules,'' {\em Nature}, vol.~467, no.~7317,
  pp.~805--810, 2010.

\bibitem{wilson2008orientation}
E.~M. Wilson-Kubalek, I.~M. Cheeseman, C.~Yoshioka, A.~Desai, and R.~A.
  Milligan, ``Orientation and structure of the ndc80 complex on the microtubule
  lattice,'' {\em The Journal of Cell Biology}, vol.~182, no.~6,
  pp.~1055--1061, 2008.

\bibitem{zhang2009dislocation}
D.-B. Zhang, R.~D. James, and T.~Dumitrică, ``Dislocation onset and nearly
  axial glide in carbon nanotubes under torsion,'' {\em The Journal of Chemical
  Physics}, vol.~130, no.~7, p.~071101, 2009.

\bibitem{XiePickering}
Z.~Xie, C.~J. Burke, B.~Mbanga, P.~T. Spicer, and T.~J. Atherton, ``Geometry
  and kinetics determine the microstructure in arrested coalescence of
  pickering emulsion droplets,'' {\em Soft Matter}, vol.~15, pp.~9587--9596,
  2019.

\bibitem{keber2014topology}
F.~C. Keber, E.~Loiseau, T.~Sanchez, S.~J. DeCamp, L.~Giomi, M.~J. Bowick,
  M.~C. Marchetti, Z.~Dogic, and A.~R. Bausch, ``Topology and dynamics of
  active nematic vesicles,'' {\em Science}, vol.~345, no.~6201, pp.~1135--1139,
  2014.

\bibitem{maroudas2020topologicalPublished}
Y.~Maroudas-Sacks, L.~Garion, L.~Shani-Zerbib, A.~Livshits, E.~Braun, and
  K.~Keren, ``Topological defects in the nematic order of actin fibres as
  organization centres of hydra morphogenesis,'' {\em Nature Physics},
  pp.~1--9, 2020.

\bibitem{Gimenez_Pinto_2017}
V.~Gimenez-Pinto, F.~Ye, B.~Mbanga, J.~V. Selinger, and R.~L.~B. Selinger,
  ``Modeling out-of-plane actuation in thin-film nematic polymer networks: From
  chiral ribbons to auto-origami boxes via twist and topology,'' {\em
  Scientific Reports}, vol.~7, mar 2017.

\bibitem{Kim2012Fabrication}
S.-H. Kim, H.~Hwang, and S.-M. Yang, ``Fabrication of robust optical fibers by
  controlling film drainage of colloids in capillaries,'' {\em Angewandte
  Chemie}, vol.~124, pp.~3661--3665, mar 2012.

\bibitem{Solomon_2018}
M.~J. Solomon, ``Tools and functions of reconfigurable colloidal assembly,''
  {\em Langmuir}, vol.~34, pp.~11205--11219, feb 2018.

\bibitem{luders2007microtubule}
J.~L{\"u}ders and T.~Stearns, ``Microtubule-organizing centres: a
  re-evaluation,'' {\em Nature Reviews Molecular Cell Biology}, vol.~8, no.~2,
  pp.~161--167, 2007.

\bibitem{artyukhov2014nanotubes}
V.~I. Artyukhov, E.~S. Penev, and B.~I. Yakobson, ``Why nanotubes grow
  chiral,'' {\em Nature Communications}, vol.~5, no.~1, pp.~1--6, 2014.

\bibitem{Head53}
A.~Head, ``X. the interaction of dislocations and boundaries,'' {\em The
  London, Edinburgh, and Dublin Philosophical Magazine and Journal of Science},
  vol.~44, no.~348, pp.~92--94, 1953.

\bibitem{mitchell1961buckling}
L.~Mitchell and A.~Head, ``The buckling of a dislocated plate,'' {\em Journal
  of the Mechanics and Physics of Solids}, vol.~9, no.~2, pp.~131--139, 1961.

\end{thebibliography}
\bibliographystyle{ieeetr}

\clearpage

\appendix
\section{Details of the discrete model}
    We consider a tubular crystal as a collection of hard spheres with positions $\mathbf{x}_i$ forming a two-dimensional triangular lattice embedded in three spatial dimensions, with a preferred lattice spacing, $a$. Neighbor-bonds, which would lie along lattice lines in a planar crystal, here lie along three families of helices.  The geometry and chirality of a pristine (defect-free) tubular crystal can be conveniently defined using a pair of integer parastichy numbers $(m,n)$, which give the number of distinct helices in the steepest right-handed and steepest left-handed families, respectively. Then the tube radius is defined by \mbox{$R=\frac{a}{2\pi}\sqrt{m^2+n^2-mn}$}. The orientation of the lattice depends on the angle between the steepest left-handed helix and the cylinder axis as $\tan \phi \approx \frac{2}{\sqrt{3}}(\frac{m}{n}-\frac{1}{2})$ (Fig.~\ref{Fig:Geometry}) \cite{BellerPRE16}.

	The discrete elastic energy of the tubular crystal is given by Eq.~\ref{Fe}. The local curvature at each node is computed as $G_i^2= 4(H_i)^2-2 K_i$, where the Gaussian curvature $K_i=(2\pi-\sum \rho_j)/A_i$ is expressed through the angles between two adjacent edges $\rho_j=\angle(l_{ij},l_{ij+1})$, and the mean curvature $H_i=||\sum_j ((\mathbf{x}_i-\mathbf{x}_j)(\cot{\psi}^1_{ij}+\cot{\psi}^2_{ij}))||/(4 A_i)$ is defined over the adjacent edges, where $\psi^{\alpha}_{ij}$ $(\alpha=1,2)$ are the two angles opposite to the edge in the two triangles sharing the edge $l_{ij}$. Here, ${A}_{i}=\sum_j \mathcal{A}_j/3$ is the observed area around a node $i$, which is calculated as the average of the areas ${\mathcal{A}}_j$ of the adjacent triangular faces. 	We choose the reference state to be a pristine triangular lattice with uniform spacing $l_{ij}=a$.  In this state the elastic energy comprises only the bending energy, which depends on the cylinder radius $R$ and length $L$ as $\mathcal{\hat{F}}_b \rightarrow \pi \kappa L/R$. 
	
    To account for nonzero spontaneous curvature, we slightly modify Eq.~\ref{Fe}. We first calculate discrete mean and Gaussian curvatures, and from these calculate the discrete principal curvatures as $k_{\{1,2\}i}=H_i\pm \sqrt{H_i^2 - K_i}$. The local curvature energy density of Eq.~\ref{Fe} is then replaced by $ \tfrac{1}{2} \kappa [(k_{1i}-H_0)^2 + k_{2i}^2]$, always choosing $k_{1i}\geq k_{2i}$.

	We assume overdamped dynamics, and the positions of spheres $\mathbf{x}_i$ change to minimize the elastic energy following the pseudo-time evolution equations $ \gamma \partial \mathbf{x}_i/\partial t= - \delta \mathcal{F}_{e} / \delta \mathbf{x}_i$. Here, $\gamma$ is the friction coefficient associated with energy dissipation during relaxation of the elastic energy. For a given bond network, we allow the system to relax to an equilibrium configuration by minimizing the elastic energy over the node positions until the reduction in energy per update step becomes smaller than $10^{-7}$.
	
    A single dislocation can be inserted by removing or adding a row of nodes to the lattice up to one end of the tube, such that the lattice remains pristine (6-coordinated)  everywhere except at the dislocation, which consists of two nodes with 5 and 7 neighbors, respectively. Such a defect in a crystalline structure can be characterized by the Burgers vector connecting the gap in the Burgers circuit around the defect, $\mathbf{b}=-\oint(\partial\mathbf{u}/\partial l)dl$, where $\mathbf{u}$ is the displacement vector.     A dislocation pair nucleation can be imposed by a single bond flip that removes a bond between neighboring nodes and replaces it with a new bond normal to it,  thus creating two pairs of nodes with 5 and 7 neighbors.  
    We validated our computational approach by reproducing the results of \cite{seung1988defects} for critical bending rigidity at which a single dislocation causes a crystalline membrane to buckle. A dislocation glide move by one lattice spacing to a neighboring node is accomplished by a similar bond flip. 
    
    We assume a separation of time scales such that the elastic energy is completely relaxed to a state of mechanical equilibrium between consecutive glide moves. The direction of gliding is chosen to decrease the total elastic energy and can be along or opposite Burgers vector, or the defect can remain at the same position if it is a stable configuration depending on interaction with other defects, the tube shape, and external stresses. 
	
	This procedure gives rise to an effective energy landscape for the dislocations on the tubular crystal, whose local minima we explore in this work. For tubes with multiple dislocations, this landscape becomes high-dimensional and difficult to minimize rigorously as each test glide move requires a full minimization of elastic energy. Therefore, we examine stability by choosing  a defect at random and performing a "trial" dislocation glide in both directions, along and opposite to the Burgers vector; we then keep the configuration with the lowest total energy. If the given position of a defect is stable, i.e.\ any glide causes an increase in the total energy, then the dislocation is excluded from the next random selection until another defect is subject to glide and the bond network changes. This procedure continues while there are defects in the lattice and any possible dislocation glide step leads to a more preferable energy state. Otherwise, if all defects have been removed or are in a stable position, the configurations is considered to be at  equilibrium. In our simulations we ignore the Peierls barrier during each glide step, assuming that thermal fluctuations enable the exploration of glide moves that we undertake quasi-deterministically. However, fully stochastic update rules or simultaneous glide might lead to other mechanically stable configurations that cannot be reached with our approach. 

\section{Continuum approximation of a freestanding tubular crystal}

The total elastic energy associated with deformations of a two-dimensional crystal approximated as a continuous elastic membrane can be written as a sum of stretching and bending contributions \cite{landau_elasticity} given by 
    	\begin{align}
    		\mathcal{F} &= \mathcal{F}_{s} + \mathcal{F}_{b}  =\frac{1}{2} \int dA \left(  \lambda u^2_{ii} + 2 \mu u^2_{ik}\right)+\frac{\kappa }{2} \int dA \left( 4 H^2 - 2 K\right),
    		\label{Eq_Fe}
    	\end{align}
	where $\lambda, \mu$ are the Lam\'e coefficients associated with the two-dimensional Young's modulus $Y=4\mu(\mu+\lambda)/(2\mu+\lambda)$ and  Poisson ratio  $\sigma = \lambda/(2\mu+\lambda)$. The strain tensor $u_{ik}$  is given in terms of the displacement field $u_j$ by $u_{ik}=\frac12 (\partial_i u_k + \partial_k u_i)$ and omits a quadratic term in $u_k$ because deformations are assumed to be small. The bending energy depends on bending modulus, $\kappa$, and local curvature that is written in the form of the Helfrich energy for membranes using the local mean curvature $H$ and Gaussian curvature $K$. The relative importance of the bending and stretching contributions to the free energy, at the size scale of the tube radius $R$, is given by the dimensionless  F{\"o}ppl-von~K{\'a}rm{\'a}n number $\gamma=YR^2/\kappa$, which is usually very large because $\kappa$ is  typically small relative to $YR^2$ \cite{lidmar2003virus}.
	
	The bending energy acts to increase the tube radius, creating  a positive azimuthal strain $u_{yy}$ and negative longitudinal strain $u_{xx}$, given at large $\gamma$  by  $u_{yy}=-u_{xx}\approx (1+\nu)/(2\gamma)$ \cite{BellerPRE16}. The optimal radius of the pristine tubular crystal  is then  $R\approx R_0(1+u_{yy})$, where $R_0$ is the tube radius prescribed by minimizing  the stretching energy alone. In the vicinity of a dislocation, deviations from a perfect cylindrical geometry are necessary to produce the expected surface buckling well known in planar 2D crystals \cite{seung1988defects}.	
	
\section{Single dislocation defect}

	\begin{figure}[t]
		\centering
			\includegraphics[width=0.45\textwidth]{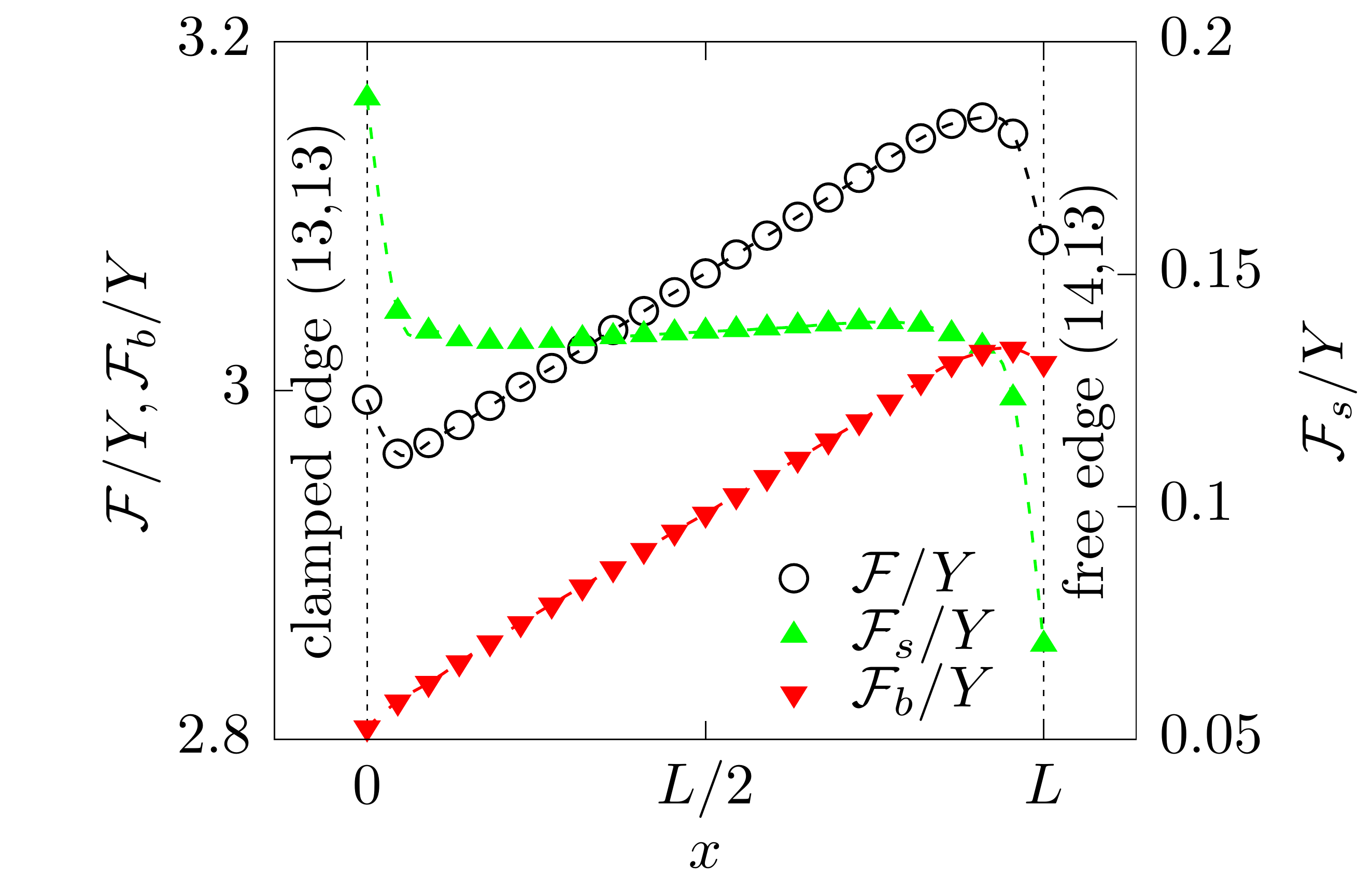} 
		\caption{ The elastic energy of a tubular crystal depending on the distance of a single dislocation from the clamped end the other, free end, with $\tilde{\kappa}=0.1, \theta=\pi/6$. In the interior interval, the constant slope in bending and total energies is defined by the energy benefit from increasing the tube radius by dislocation glide, and is proportional to the resulting change in tube curvature.} 
		\label{single}
	\end{figure}

    In the main text we ignore boundary effects that will arise when the dislocation approaches the ends of the finite-length tubular crystal. Here we justify this simplification by showing  that, for individual dislocations, the length scale for interactions between dislocations and the tube ends is very small, so that the ends have essentially no influence in the interior. 
    
    We construct an achiral lattice at $(m,n)=(13,13)$, $a=1,~\tilde{\kappa}=0.1$ with one free edge and another firmly clamped edge that at which displacement of boundary nodes is disallowed. Then we impose a single dislocation with  Burgers vector at angle $\theta=\pi/6$ to the tube axis. The dislocation causes a transition to $(m,n)=(14,13)$ in the tessellation in the part of the lattice closer to the free edge, slightly increasing the tube radius. If the tube is flexible and the dislocation is far from the tube edges, meaning that interaction with the boundaries is weak, the defect tends to glide toward the clamped edge, either along or opposite the Burgers vector, to increase the tube radius $R$ so that the bending energy is decreased. However, when the dislocation is imposed close to the free edge, at a distance less than three lattice spacings, it causes strong deformations of the tube, and the defect glides to the free edge to relieve the stretching energy, even though this bears a cost in bending energy (Fig.~\ref{single}). On the other hand, if a dislocation is very close to the clamped end of the tube, the large cost in stretching energy acts to repel the defect to a stable state at a distance of two lattice spacing from the edge (the minimum of the total energy in Fig.~\ref{single}). Such a configuration with one clamped and one free end resembles the microtubule (MT) structures with only one end firmly anchored to the organizing center where the MT starts to grow \cite{luders2007microtubule}, or assembling single-walled carbon nanotubes with a rigid contact interface between the nanotube edge and the metal catalyst where the they nucleate \cite{artyukhov2014nanotubes}. Also, we note that the numerical result for a flexible tube is similar to an interaction of a dislocation with a grain boundary in a 2D lattice \cite{Head53}, but only at small distance from the boundaries, where the forces exerted by the interface dominate and a dislocation is attracted to the free surface and repelled from an interface of larger shear modulus. 
	
	The energy savings due to the increasing tube radius $R$ when a dislocation glides along the tube defines the constant slope in bending energy with respect to glide distance for defects far from boundaries. Assuming that the transition zone, the region of a change in the parastichy numbers and tessellation, is narrow  in the vicinity of a dislocation, the approximation for the slope is $\pi \kappa (1/R_1 - 1/R_2)$, where $R_1,R_2$ are the tube radii of different tubular tessellations. The stretching energy in Fig.~\ref{single} is an order of magnitude smaller than bending energy because a dislocation causes a local in-plane stress whereas the lattice has unavoidable bending energy at each node if there is no spontaneous curvature. 
		
	By this preliminary example we demonstrate that dislocations interact with the boundary via buckling and in-surface stresses. The direction of glide depends on the initial position and the distance from the boundary, and can be changed by the boundary conditions. This also allows us to estimate the deformation length around a defect at which the defect starts to interact significantly with the boundary. The deformation length caused by a dislocation in a tubular crystal is smaller than for a circular plate  due to the cylindrical geometry constraint; for an isolated dislocation in a lattice of circular shape with free boundaries, the length was estimated to be of order $10a$ at $\tilde{\kappa}=0.1$ \cite{mitchell1961buckling,seung1988defects}. Thus, imposing defects farther than this deformation length from the boundaries allows us to avoid boundary effects.     
		
    \section{Interacting dislocations on a rigid cylindrical substrate}	

	\begin{figure*}[t]
		\centering
		\begin{tabular}{cc}
			(a)&(b)\\
			\includegraphics[width=0.45\textwidth]{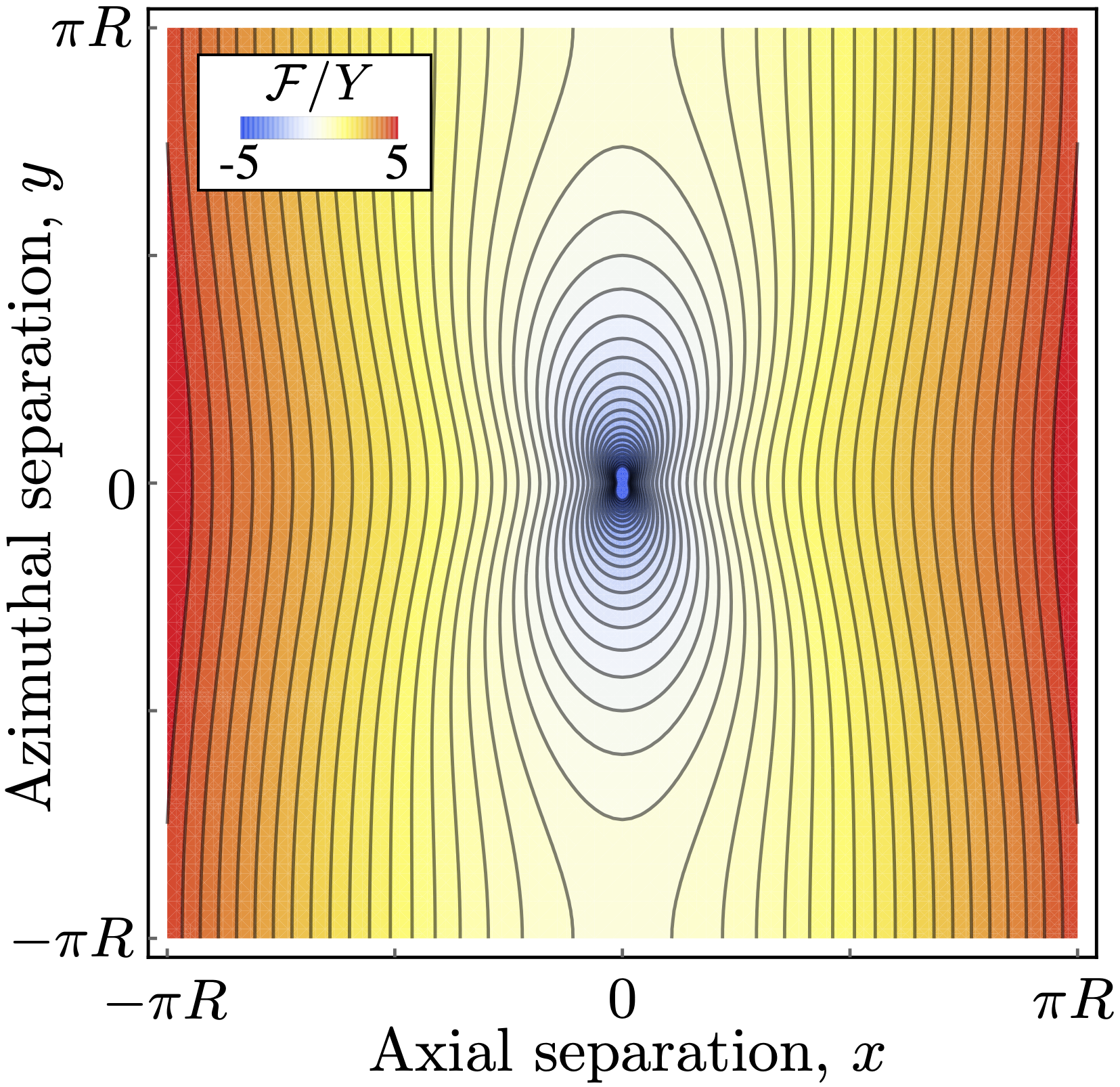} 
		   &\includegraphics[width=0.45\textwidth]{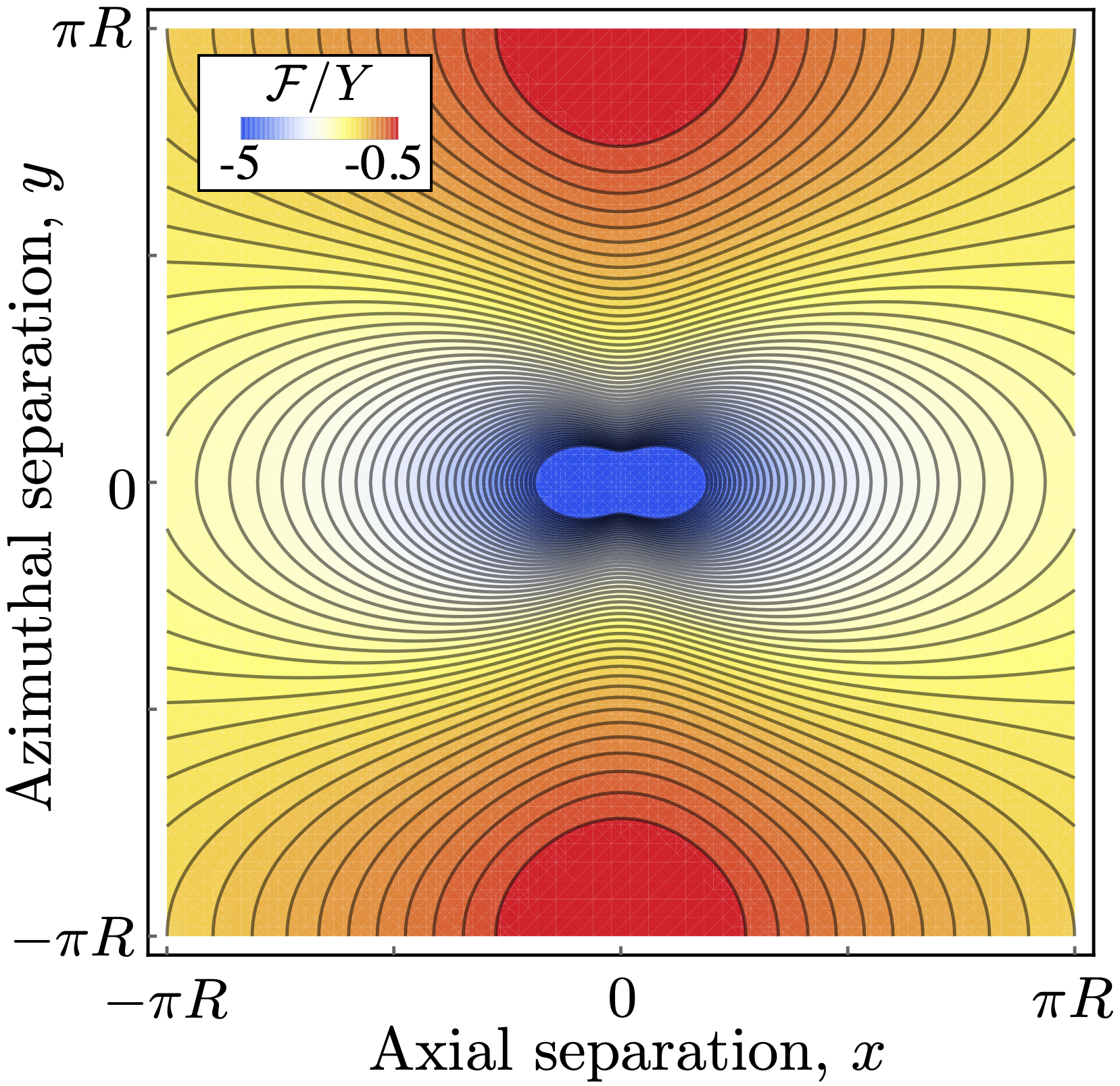}
		\end{tabular}
		\caption{Energy landscape for interaction energy of two antiparallel dislocations gliding along (a) the azimuthal direction at $\theta=\pi/2$ and (b) along the tube axis at $\theta=0$ calculated according to Eq.~4 in the main text. }
		\label{fig:Amir}
	\end{figure*}

    Here, for completeness, we provide stability diagrams for the dislocation pair energy on a cylinder calculated using Eq.~4 in the main text proposed in \cite{Amir13,BellerPRE16}. In Fig.~\ref{fig:Amir} we plot the interaction energy corresponding to two separate dislocations with antiparallel Burgers vectors allowed to glide in the azimuthal ($\theta=\pi/2$) and longitudinal ($\theta=0$) directions. The energy landscapes are symmetric about the $x$ and $y$ axes, in contrast to our results for the  freestanding tubular crystal, which is only symmetric under $y\rightarrow -y$, due in part to the bending energy. At $\theta=\pi/2$ there is an unstable equilibrium (saddle point) at $x=0, y=\pi R$  separation and two identical stable states at $x \neq 0$ with the separation vector at $45^{\circ}$ to the $x$ axis.  In the case of $\theta=0$, a single maximum exists at $x=0, y \neq 0$ separating two symmetric stable states with similar separation vector at $45^{\circ}$ to the $x$ axis. The absolute energy minimum is at the origin where the two dislocations annihilate, leaving behind a pristine lattice.  
    
    \section{Computation of the local tube radius}
    
    The local radius $R_i$ at a node $i$ is calculated as the shortest distance between the node and the centerline of the tube, $C$. The centerline can, in general, significantly deviate from the axis in a pristine tube and have kinks along it due to dislocations. For tubes with a small deviation of the tube axis from the coordinate axis $OX$,  we calculate the position of the centerline at discrete points lying in evenly separated parallel cross sections of the tube with the normal vector along $OX$. Position $C(y,z)$ at each fixed $x$ is found as the equilibrium position where the $y$ and $z$ components of a fictitious repulsive force, exerted by all nodes in the lattice, vanish. The force from each node $R_j$ is oriented toward $C(y,z)$ and has a power-law form, with amplitude decaying with the distance to a node as $|R_j-C(y,z)|^{-8}$. The separation between cross sections is chosen to be small ($0.01a$) allowing a dense discretization of the centerline. Then the local radius $R_i$ is computed as the distance to the nearest point on $C(x,y,z)$. In case of large deviations of the tube axis from $OX$, we calculate $C$ in cross sections constructed in a local frame rotated such that the local $OX$ coincides with the tangent of $C$ calculated at previous step. This ensures small separation between points along $C$.
    
\end{document}